\documentclass[prd,aps,a4paper,eqsecnum,floatfix,nofootinbib,twocolumn]{revtex4}  %

\usepackage{graphicx} 
\usepackage{amsmath,amsfonts,amssymb}
\usepackage{tikz}
\usepackage{appendix}
\usepackage[caption=false,justification=justified]{subfig}

\newcommand{\beq}{\begin{equation}}
\newcommand{\eeq}{\end{equation}}
\newcommand{\bea}{\begin{eqnarray}}
\newcommand{\eea}{\end{eqnarray}}
\newcommand{\be}{\begin{equation}}      
\newcommand{\ee}{\end{equation}}

\def\nn{\nonumber}

\begin{document}

\title{Scalar waves from unbound orbits in a TS spacetime: PN reconstruction of the field and radiation losses in a  self-force approach}

\author{Giorgio Di Russo$^{1}$, Massimo Bianchi$^{2}$, Donato Bini$^{3}$}
  \affiliation{
$^1$School of Fundamental Physics and Mathematical Sciences, Hangzhou Institute for Advanced Study, UCAS, Hangzhou 310024, China\\
$^2$Dipartimento di Fisica, Universit\`a di Roma \lq\lq Tor Vergata" and Sezione INFN Roma2, Via della
Ricerca Scientifica 1, 00133, Roma, Italy\\
$^3$Istituto per le Applicazioni del Calcolo ``M. Picone,'' CNR, I-00185 Rome, Italy
}

\date{\today}

\begin{abstract}
We analyze scalar wave emission from unbound orbits in a Topological Star spacetime. Our study uses a  self-force approach and leads to a  
Post-Newtonian reconstruction of the field  along the orbit, both in the time domain and in the frequency domain.  We also compute leading-order radiation losses, namely energy and angular momentum.
\end{abstract}

\maketitle

\section{Introduction}
In the recent years gravitational self-force (GSF) has been studied successfully in Schwarzschild and Kerr black hole (BH) spacetimes (see e.g.,   \cite{Poisson:2011nh,Barack:2018yvs,Bini:2013zaa,Kavanagh:2015lva,Shah:2012gu,Bini:2015xua} and references therein). 
 A test particle perturbing these spacetimes  
realizes a gravitational two-body system, i.e., the cornerstone  on which the study of the gravitational interaction itself lies. The GSF approach is now standard in BH spacetimes: after writing the perturbation equations one first decouples the angular variables by expanding the perturbation in spherical (or spheroidal in the Kerr case) harmonics, then Fourier-transforms the time variable, and finally is left with a \lq\lq main" radial equation. The latter can be solved perturbatively in various approximation schemes: Post-Newtonian (PN) expansion, implying a weak field and slow motion (see, e.g., \cite{Blanchet:2013haa}), Post-Minkowskian (PM) expansion, implying a weak field but allowing for relativistic motions (see, e.g., \cite{Damour:2016gwp}), Jeffreys-Wentzel-Kramers-Brillouin (JWKB) semi-classical approximation (large $l$ quantum numbers), Mano-Suzuki-Tagasugi (MST) expansion in (confluent) hypergeometric functions, satisfying the correct boundary conditions of purely ingoing waves at the horizon and outgoing waves at infinity (see, e.g., \cite{Mano:1996vt,Sasaki:2003xr}).
Writing the solution to the radial equation according to these schemes can be anyway performed in PN sense, i.e., limiting the solution at some, specified PN order.
Other complications arise when trying to reconstruct the perturbing field, because one has to subtract the contribution of the singular field (associated with the particle's world line where its mass-energy distribution is concentrated) before summing up over the harmonics. Furthermore, to reach a high-PN result  for the reconstructed perturbing field or for the evaluation of any associated gauge-invariant quantity one has to combine these methods, e.g., PN and MST, whereas the analytic reconstruction of the singular field can be eventually accomplished by using the JWKB approach.
All these steps have been standardized over the years and are discussed in several papers (see, e.g., Refs. \cite{Bini:2013zaa,Bini:2013rfa,Bini:2014nfa,Bini:2015bla,Kavanagh:2015lva,Bini:2015bfb,Kavanagh:2016idg} for the pioneering analytical self-force computations in the Schwarzschild spacetime for both circular and quasi circular orbits of the source).

GSF applications in Schwarzschild and Kerr spacetimes led to a number of results concerning circular and quasi-circular bound orbits (expanded in small-eccentricity).
Only last year scalar waves from unbound orbits in a Schwarzschild spacetime have been studied \cite{Bini:2024icd}. The main difficulty in this case is the continuous frequency spectrum, whereas for circular and quasi-circular bound orbits (i.e., eccentric orbits expanded in powers of the eccentricity  up to a  certain power) one  always has a countable number of frequencies. In addition,  passing to the frequency domain requires the evaluation of nontrivial integrals, involving Bessel (and iterated Bessel) functions.

In Ref. \cite{Bianchi:2024vmi} the MST approach has been generalized outside the BH context. More precisely, the MST technology has been implemented for a topological star (TS) whose metric is smooth and horizon-free in $D=5$, at least in a certain region of the parameter space, and displays a naked singularity when reduced to $D=4$. In that case  scalar waves from circular orbits have been studied analytically \cite{Bianchi:2024rod}.
Moreover the novel approach based on the quantum Seiberg-Witten curves for $N=2$ supersymmetric quiver gauge theories \cite{Aminov:2020yma, Bianchi:2021xpr, Bonelli:2021uvf, Bianchi:2021mft, Fioravanti:2021dce, Bianchi:2022qph, Consoli:2022eey, Bonelli:2022ten, Gregori:2022xks, 
Fucito:2023afe, Bautista:2023sdf, Bianchi:2023rlt, Aminov:2023jve, DiRusso:2024hmd, Fucito:2024wlg, Bianchi:2024mlq, Cipriani:2024ygw, Bena:2024hoh, Dima:2024cok, Dima:2025zot,Bianchi:2021yqs,Bianchi:2022wku,Cipriani:2025ikx} and on the AGT correspondence with 2-d CFT's \cite{Alday:2009aq} naturally produced the double PM-PN expansion \cite{Fucito:2023afe, Bautista:2023sdf, Bianchi:2023rlt, Aminov:2023jve, DiRusso:2024hmd, Fucito:2024wlg}, that we will look in future applications. 

With the present work we proceed further with our study, examining the case of unbound orbits.
Unfortunately (but expectedly) new computational difficulties arise in this case, and one is left with very long (and memory consuming) calculations which forbid, at least for the moment, expanding to a very high PN order as in the BH case.
 
We will limit our discussion to the full reconstruction of the field, either in the time domain or in the frequency domain, as well as to the (leading-order) energy and angular momentum loss, learning from the present study the underlying difficulties for this kind of computations and paving the way for future works, where gauge-invariant quantities like the scattering angle can be tackled.

In order to be prepared for a  broader  discussion, involving different metrics, let us start by considering general  static and spherically symmetric metrics in four and five spacetime dimensions of the type
\bea
\label{met_gen_d}
ds^2&=&-f_t(r)dt^2+f_r(r)dr^2 +f_{\Omega}(r)(d\theta^2+\sin^2\theta d\phi^2)\nonumber\\
&+& (D-4)f_y(r)dy^2\,,
\eea 
with coordinates $(t,r,\theta,\phi,y)$ adapted to the Killing symmetries of the spacetime, and corresponding to non-vacuum solutions. 
Here $D$ denotes the spacetime dimension: $D=4$ as usual for BHs while $D=5$ in the case of TS spacetimes.
Sometimes, to ease comparisons  with other 4D spacetimes we will consider the reduction to $D=4$ of the TS metric, i.e., a 4D spacetime with metric equal to the one induced on the leaves $y=$constant.

\section{Scalar self-force analysis}

Let us  consider the scalar wave equation in the  general situation \eqref{met_gen_d}, which later we specialize  to the TS case,
\beq
\label{eq_fund}
\Box \psi=-4\pi \rho\,,
\eeq
where
\beq
\Box \psi=\frac{1}{\sqrt{-g}}\partial_\mu (\sqrt{-g}g^{\mu\nu}\partial_\nu  \psi)\,,
\eeq
in a system of spherical-like coordinates $(t,r,\theta,\phi, {\mathbf y})$, with in general ${\mathbf y}=(y_1, \ldots, y_{D-4})$, and scalar charge density given by
\bea
\rho(t,r,\theta,\phi,{\mathbf y})&=&q\int \frac{d\tau}{\sqrt{-g}} \delta^{(D)}(x-x_p(\tau))\nonumber\\
&=& \frac{q}{u^t({t})\sqrt{-g}}\delta(r-r_p(t))\delta(\theta-\frac{\pi}{2})\times \nonumber\\
&\times & \delta(\phi-\phi_p(t))\delta(y-y_p(t))\ldots\,,
\eea
where $y_1=y$ (when $D=5$) to ease notation, $x=x_p^\alpha(\tau)$ denote the (parametric) equation of the timelike world line of the massive particle emitting the scalar waves. Without loss of generality we consider equatorial motion, namely
\beq
t=t_p(\tau)\,,\quad r=r_p(\tau)\,,\quad \theta_p(\tau)=\frac{\pi}{2}\,,\quad \phi=\phi_p(\tau) \,.
\eeq
The associated four-velocity is denoted by $u^\alpha={dx_p^\alpha}/{d\tau}\big|_{\tau=\tau_p(t)}$ so that $u^t={dt_p}/{d\tau}\big|_{\tau=\tau_p(t)}$.

Let us specialize the metric \eqref{met_gen_d} to $D=4$, and use the identity
\bea
\delta(\theta-\frac{\pi}{2})\delta(\phi-\phi_p(t))&=&\sum_{lm}Y_{lm}(\theta,\phi)Y_{lm}^*(\frac{\pi}{2},\phi_p(t))\nonumber\\
&=& \sum_{lm}e^{-im\phi_p(t)}Y_{lm}(\theta,\phi)Y_{lm}^*(\frac{\pi}{2},0)\,.\nonumber\\
\eea
We find
\beq
\sqrt{-g}=\sqrt{f_t(r) f_r(r)}\, f_\Omega(r)\sin(\theta)\,,
\eeq
which reduces to $r^2\sin\theta$ in the Schwarzschild case, i.e., for $f_r(r)=\frac{1}{f_t(r)}$ and $f_\Omega(r)=r^2$.
The inhomogeneous scalar wave equation, for an ansatz of the form
\beq
\label{psi_deco}
\psi(t,r,\theta,\phi)=\sum_{lm}Y_{lm}(\theta,\phi)\int \frac{d\omega}{2\pi}e^{-i\omega t} R_{lm\omega} (r)\,,
\eeq
reads 
\begin{widetext}
\bea
\label{eq_inhom}
&&\sum_{lm}Y_{lm}(\theta,\phi)\int \frac{d\omega}{2\pi}e^{-i\omega t}\frac{1}{f_r}{\mathcal L}_r R_{lm\omega}=-4\pi \sum_{lm}Y_{lm}(\theta,\phi)\int \frac{d\omega}{2\pi}e^{-i\omega t} \hat T_{lm\omega}(r) \,,
\eea
where
\beq
{\mathcal L}_r R_{lm\omega}=  \frac{d^2}{dr^2}R_{lm\omega}+ \left(\frac{f_\Omega'}{f_\Omega} -\frac{f_r'}{2f_r} +\frac{f_t'}{2f_t}\right)\frac{d}{dr}R_{lm\omega}
+f_r \left(\frac{\omega^2}{f_t}-\frac{L}{f_\Omega}\right)R_{lm\omega}\,,
\eeq
with $L=l(l+1)$ and
\beq
\hat T_{lm\omega}(r)=q Y_{lm}^*(\frac{\pi}{2},0)\int dt e^{i\omega t}\frac{e^{-im\phi_p(t)}}{u^t({t})\sqrt{f_t(r) f_r(r)}\, f_\Omega(r)}\delta(r-r_p(t))\,.
\eeq
\end{widetext}
By using the operator ${\mathcal L}_r$ Eq. \eqref{eq_inhom} implies
\beq
\label{inhom_rad_eq}
{\mathcal L}_r R_{lm\omega}=S_{lm\omega}(r)\,,
\eeq
where
\bea
S_{lm\omega}(r)&=& -4\pi f_r(r) \hat T_{lm\omega}(r)\nonumber\\
&=& -4\pi q  Y_{lm}^*(\frac{\pi}{2},0) {\mathcal F}_{lm\omega}(r)\,,
\eea
and
\bea
\label{integ_cal_F}
{\mathcal F}_{lm\omega}(r)= \int dt \frac{e^{i(\omega t-m\phi_p(t))}\sqrt{f_r(r)} \delta(r-r_p(t))}{u^t({t})\sqrt{f_t(r)}\, f_\Omega(r)}\,.\qquad
\eea
Eq. \eqref{inhom_rad_eq} is solved by using the Green function method, i.e., introducing the Green function $G_{l{}\omega}(r,r')$ such that
\bea
{\mathcal L}_r G_{l{}\omega}(r,r')=\frac{f_r(r)}{f_\Omega(r)}\delta(r-r')\,,
\eea
implying
\beq
R_{lm\omega} =\int dr'  \frac{f_\Omega(r')}{f_r(r')} G_{l{}\omega}(r,r')S_{lm\omega}(r')\,.
\eeq
Notice that \cite{Bini:2013zaa,Bini:2013rfa}
\bea
G_{l{}\omega}(r,r')&=&\frac{R_{\rm in}^{l{}\omega}(r)R_{\rm up}^{l{}\omega}(r')}{W_{l{}\omega}} H(r'-r)\nonumber\\
&+&
\frac{R_{\rm in}^{l{}\omega}(r')R_{\rm up}^{l{}\omega}(r)}{W_{l{}\omega}} H(r-r')\nonumber\\
&\equiv & G_{l{}\omega}^-(r,r') H(r'-r)\nonumber\\
&+& G_{l{}\omega}^+(r,r') H(r-r')
\,,
\eea
where $H(x)$ denotes the Heaviside step function, is $m$-independent in the spherically symmetric case since $R_{\rm in}^{l{}\omega}(r)$ and $R_{\rm up}^{l{}\omega}(r)$, the two independent solutions of the homogeneous radial equation (with proper boundary conditions), are  $m$-independent, too. Similarly their constant Wronskian
\beq
W_{l{}\omega}=\frac{f_\Omega(r)}{f_r(r)} [R_{\rm in}^{l{}\omega}(r) R_{\rm up}'{}^{l\omega}(r)-R_{\rm in}'{}^{l\omega}(r) R_{\rm up}^{l\omega}(r)]\,,
\eeq
is $m$-independent too. Note that one has either a `left' or a `right' Green function $G^\pm_{l\omega}(r,r')$,
namely,
\bea
G_{l{}\omega}^-(r,r')&=&\frac{R_{\rm in}^{l{}\omega}(r)R_{\rm up}^{l{}\omega}(r')}{W_{l{}\omega}}\,,\nonumber\\
G_{l{}\omega}^+(r,r')&=&  \frac{R_{\rm in}^{l{}\omega}(r')R_{\rm up}^{l{}\omega}(r)}{W_{l{}\omega}}\,,
\eea
according to the relative radial positions $r<r'$ or $r>r'$. This will imply a corresponding  left and right scalar field. However, when evaluating the field at the particle's location for self-force purposes, the field (but not its derivatives) is necessarily continuous and can be computed (as we do below) with any one of these two fields.
Finally, Eq. \eqref{psi_deco} becomes
\begin{widetext}
\bea
\label{psi_val}
\psi(t,r,\theta,\phi)
&=&-4\pi q \sum_{lm}Y_{lm}(\theta,\phi)Y_{lm}^*(\frac{\pi}{2},0)\int \frac{d\omega}{2\pi}e^{-i\omega t} \int dr'  \frac{f_\Omega(r')}{f_r(r')} G_{l{}\omega}(r,r') {\mathcal F}_{lm\omega}(r') \nonumber\\
&=&-4\pi q \sum_{lm}Y_{lm}(\theta,\phi)Y_{lm}^*(\frac{\pi}{2},0)\int \frac{d\omega}{2\pi}e^{-i\omega t} \int dr' G_{l{}\omega}(r,r')
\int dt'   \frac{e^{i(\omega t'-m\phi_p(t'))}  \delta(r'-r_p(t'))}{u^t({t'})\sqrt{f_t(r')f_r(r')}}\nonumber\\
&=&-4\pi q \sum_{lm}Y_{lm}(\theta,\phi)Y_{lm}^*(\frac{\pi}{2},0)\int dt'  \int \frac{d\omega}{2\pi}e^{-i\omega t}   G_{l{}\omega}(r,r_p(t'))
 \frac{e^{i(\omega t'-m\phi_p(t'))}  }{u^t({t'})\sqrt{f_t(r_p(t'))f_r(r_p(t'))}}\,.
\eea
Eq. \eqref{psi_val} identifies then
\bea
\label{R_pm}
R_{lm\omega}^\pm (r)=-4\pi q Y^*_{lm}(\frac{\pi}{2},0) \int dt' G_{l{}\omega}^\pm (r,r_p(t'))
 \frac{e^{i(\omega t'-m\phi_p(t'))}  }{u^t({t'})\sqrt{f_t(r_p(t'))f_r(r_p(t'))}}\,.
\eea

We will be interested in evaluating the field along the particle's world line, namely
\bea
\psi(t,r_p(t),\frac{\pi}{2},\phi_p(t))
&=&-4\pi q \sum_{lm}|Y_{lm}(\frac{\pi}{2},0)|^2\int dt'  \int \frac{d\omega}{2\pi}e^{-i(\omega t-m\phi_p(t))}   G_{l{}\omega}(r_p(t),r_p(t'))
 \frac{e^{i(\omega t'-m\phi_p(t'))}  }{u^t({t'})\sqrt{f_t(r_p(t'))f_r(r_p(t'))}}\,,\nonumber\\
\eea
as well as in evaluating its Fourier transform (FT)
\bea
\hat\psi (\omega)=\int dt e^{i\omega t}\psi(t,r_p(t),\frac{\pi}{2},\phi_p(t))\,,
\eea
together with the \lq\lq soft limit" $\langle \psi \rangle=\lim_{\omega \to 0} \hat\psi (\omega)$, i.e. the time average (whence the notation), that yields
\bea
\langle \psi \rangle=\int dt\, \psi(t,r_p(t),\frac{\pi}{2},\phi_p(t))\,.
\eea
Notice that Eq. \eqref{psi_val} looks much simpler in the case of a circular orbit at $r_p(t)=r_0$  and $\phi_p(t)=\Omega t$ (with $u^t = u^t_0$ independent of $t$, too) whereby
\bea
\label{psi_val_circ}
\psi(t,r,\theta,\phi)
&=&-4\pi q \sum_{lm}Y_{lm}(\theta,\phi)Y_{lm}^*(\frac{\pi}{2},0)\int dt'  \int \frac{d\omega}{2\pi}e^{-i\omega t}   G_{l{}\omega}(r,r_0)
 \frac{e^{i(\omega -m\Omega) t'}  }{u^t_0\sqrt{f_t(r_0)f_r(r_0)}}\nonumber\\
&=&
-4\pi q \sum_{lm}Y_{lm}(\theta,\phi)Y_{lm}^*(\frac{\pi}{2},0) \int \frac{d\omega}{2\pi}e^{-i\omega t}   G_{l{}\omega}(r,r_0)
 \frac{2\pi \delta(\omega -m\Omega)  }{u^t_0\sqrt{f_t(r_0)f_r(r_0)}}\nonumber\\
&=&
-4\pi q \sum_{lm}Y_{lm}(\theta,0)Y_{lm}^*(\frac{\pi}{2},0)  e^{im(\phi-\Omega t)}   
 \frac{G_{l{}\omega}(r,r_0)|_{\omega=m\Omega}}{u^t_0\sqrt{f_t(r_0)f_r(r_0)}}\,.
\eea
Along the particle's world line we have $\phi=\Omega t$ and hence
\bea
\psi(t,r_0,\frac{\pi}{2},\Omega t)
&=&-4\pi q \sum_{lm}|Y_{lm}(\frac{\pi}{2},0)|^2  
 \frac{G_{l{}\omega}(r_0,r_0)|_{\omega=m\Omega}}{u^t_0\sqrt{f_t(r_0)f_r(r_0)}}\,,
\eea
which actually does not depend  on $t$, since the dependence will be on $\phi-\Omega t$ which cancels along the orbit.

\end{widetext}

\section{Top Star metric} 

A TS is described by the following $D=5$ metric \cite{Bah:2020ogh}
\bea\label{metric}
ds^2&=&-f_s(r)dt^2+\frac{dr^2}{f_s(r)f_b(r)}\nn\\
&+& r^2(d\theta^2+\sin^2\theta d\phi^2)+f_b(r)dy^2\,,
\eea
with 
\beq
f_{s,b}(r)=1-\frac{r_{s,b}}{r}.
\eeq
Comparing with Eq. \eqref{met_gen_d} we identify then
\beq
f_t(r)=f_s(r)\,,\quad f_r(r)=\frac{1}{f_s(r)f_b(r)}\,,\quad f_\Omega(r)=r^2\,.
\eeq
The coordinate $y$ is compact $y\sim y+2\pi R_y$. The metric \eqref{metric} is a magnetically charged solution of $D=5$ Einstein-Maxwell system of equations~\footnote{The electric solution corresponds to a string wound around the compact $y$ direction.}, where the electromagnetic field is given by
\beq
A=-P\cos\theta d\phi,\,,\qquad F=P\sin\theta d\theta\wedge d\phi\,,
\eeq
and with 
\beq
P^2=\frac{3r_b r_s}{2\kappa_5^2}\,,
\eeq
representing the magnetic charge.
Actually, with this form of the metric there exist two different regimes: 1) black string (BS) regime: $r_b<r_s$, with an event horizon at $r=r_s$; 2) TS regime: $r_s<r_b$, a smooth
horizon-less solution if $R_y=2r_b^{3/2}/\sqrt{r_b-r_s}$, see e.g.  \cite{Bah:2020ogh}. Furthermore, stability against (metric) linear perturbations requires $r_b{<}r_s{<}2r_b$
in the BS case and $r_s{<}r_b{<}2r_s$ in the TS one (see e.g., Refs. \cite{Bianchi:2023sfs, DiRusso:2024hmd, Cipriani:2024ygw,Bena:2024hoh,Dima:2024cok,Dima:2025zot}).

After dimensional reduction to $D = 4$, the solution exposes a naked singularity
and corresponds to a massive source with mass
\beq
G_4M_{\rm TS} = \frac{r_s}2  + \frac{r_b}4 \,, 
\eeq
with
\beq
8\pi G_4=\kappa_4^2=\frac{\kappa_5^2}{2\pi R_y}=\frac{8\pi G_5}{2\pi R_y}\,.
\eeq
Hereafter we will set $G_4\equiv G = 1 = c$. For $r_b = 0$, and
thus $r_s = 2GM_{\rm TS}$, the resulting singular solution is 
Schwarzschild BH$\times S_1$.
Often we will find it convenient to denote
$r_s=2M$, $r_b=\alpha r_s=2\alpha M$,
with $M$ a common length scale, not to be confuse with $M_{\rm TS}=\frac{r_s}{4}(\alpha+2)$, the mass of the TS, unless for $r_b=0$.
 
In the TS case, $r_s{<}r_b{<}2r_s$, the geometry smoothly ends with a `cap' at $r=r_b$, and the coordinate singularity at $r=r_s$ disappears together with the curvature singularity at $r=0$. Indeed, setting $\tilde{r} = \sqrt{r_b(r-r_b)}$ and $\tilde{\varphi} = y/R_y$, the near-cap (spatial) geometry is simply the product of a 2-sphere of radius $r_b$ in the $\theta$ and $\phi$ directions times a 2-plane  in the $\tilde{r}$ and $\tilde{\varphi}$ direction. Meanwhile the red-shift factor takes the constant value $0<f_s(r_b) = (r_b-r_s)/r_b<1/2$.

The surfaces  at $r=$constant correspond to null surfaces, since the 1-form $dr$ has a magnitude
\beq
dr\cdot dr =g^{rr}=f_s(r)f_b(r)\,,
\eeq
which vanishes at $r=r_s$ and $r=r_b$. However, the latter value is the boundary of the allowed region, and it should be considered only in the sense of the limit $r\to r_b^+$.

Therefore, even if one may consider the TS metric as a minimal modification to the Schwarzschild (or Reissner-Nordstr\"om) spacetime (one additional coordinate, $y$, and one parameter $r_b$, non-vacuum) the geometrical as well as the physical content is significantly different.

\section{self-force and unbound motions}
 
For a TS spacetime with metric  induced on the $y=$ constant leaves (or, equivalently the metric \eqref{met_gen_d} in the case $D=4$), Eq. \eqref{psi_val} implies
\begin{widetext}
\bea
\label{psi_val_TS}
\psi(t,r,\theta,\phi)
&=&-4\pi q \sum_{lm}Y_{lm}(\theta,\phi)Y_{lm}^*(\frac{\pi}{2},0)\int dt'  \int \frac{d\omega}{2\pi}e^{-i\omega t}   G_{l{}\omega}(r,r_p(t'))
 \frac{e^{i(\omega t'-m\phi_p(t'))} \sqrt{f_b(r_p(t'))} }{u^t(t')} \,,
\eea
and in general along the orbit (assumed to be equatorial, $\theta_p(\tau)=\frac{\pi}{2}$, without loss of generality both in the geodetic and in the non-geodetic case)
\bea
\label{psi_val_TS_orbit}
\psi(t,r_p(t),\frac{\pi}{2},\phi_p(t))
&=&-4\pi q \sum_{lm}|Y_{lm}(\frac{\pi}{2},0)|^2  \int dt'  {\mathcal A}_{lm}(t,t')\,,
\eea
with~\footnote{Let us recall the physical dimensions $[{\mathcal A}_{lm}]\sim \frac{1}{L^2}$, 
$ \psi\sim  q/L=$ dimensionless.
}
\bea
\label{main_integ}
{\mathcal A}_{lm}(t,t')&=&e^{i m (\phi_p(t)-\phi_p(t'))}\frac{\sqrt{f_b(r_p(t')} }{u^t(t')} \int \frac{d\omega}{2\pi} e^{-i\omega (t-t')}    G^-_{l{}\omega}(r_p(t),r_p(t'))\nonumber\\
&=&e^{i m (\phi_p(t)-\phi_p(t'))}\frac{\sqrt{f_b(r_p(t')} }{u^t(t')} \int \frac{d\omega}{2\pi} e^{-i\omega (t-t')}  \frac{R_{\rm in}^{l{}\omega}(r_p(t))R_{\rm up}^{l{}\omega}(r_p(t'))}{W_{l{}\omega}}\nonumber\\
&\equiv &  A^{m}_1(t,t') \int \frac{d\omega}{2\pi} e^{-i\omega (t-t')}A^{l\omega}_2(t,t')  \,,
\eea
\end{widetext}
where
\bea
A^{m}_1(t,t')&=& e^{i m (\phi_p(t)-\phi_p(t'))}\frac{\sqrt{f_b(r_p(t')} }{u^t(t')}\,,\nonumber\\
A^{l\omega}_2(t,t')&=& \frac{R_{\rm in}^{l{}\omega}(r_p(t))R_{\rm up}^{l{}\omega}(r_p(t'))}{W_{l{}\omega}}\,,
\eea
and where we used the left Green function 
\beq
G_{l{}\omega}^-(r,r')=\frac{R_{\rm in}^{l{}\omega}(r)R_{\rm up}^{l{}\omega}(r')}{W_{l{}\omega}}\,,
\eeq
thanks to the continuity of the field at the particle position.

To proceed further we need to specify the orbit of the source. We will consider both the cases of 1) spatial straight line motions (which is nevertheless spacetime accelerated, and hence not as simple as in the flat space analogous situation) 
and 2) geodesic unbound  motion  in the equatorial plane. 
Our task will be here the full reconstruction of the scalar field along the source's world line in PN sense. Indeed,  we will show the first few PN terms only, being limited by the long computational time needed to reach a high PN accuracy level. This work, however, will display all the main underlying technical difficulties which we will overcome here, leaving the task of going beyond the present analysis in the future, when either a more efficient manipulation of such expressions or a new generation of computers together with a more powerful version of algebraic manipulator systems will be available.

\subsection{A warming up exercise: spatial straight line motion}

To explore a simplified situation we limit our considerations to spatial straight line motion (but spacetime curved motion, i.e., accelerated motion).
The latter choice has advantages (it is simple) and disadvantages (it is clearly non-geodetic, and approaches the geodetic behavior only asymptotically). Actually, the problem of radiation emission in this context has been considered long ago by Ruffini and Wheeler \cite{rr1,rr2} and bears the the name of \lq\lq splash radiation."
 
The (spatial) parametric equations of the particle's orbit (a straight line in spherical coordinates flat spacetime) are then given by
\beq
r_p(t)=\sqrt{b^2+v^2t^2}\,,\qquad \phi_p(t)={\rm arctan}\left(\frac{vt}{b}\right)\,,
\eeq
using the coordinate time as a parameter along the orbit and with 
\beq
\frac{dr_p(t)}{dt}=\frac{tv^2}{r_p(t)}\,,\quad  \frac{d\phi_p(t)}{dt}=\,\frac{bv}{r_p(t)^2}.
\eeq
 Note that passing to standard Cartesian coordinates in flat spacetime,  $x=r\cos\phi$, $y=r\sin\phi$, the above equations reduce to
\beq
x=b\,,\qquad y=vt\,.
\eeq
In the TS spacetime one can identify the associated four velocity vector $u^\alpha={dx_p^\alpha}/{d\tau}$ of the orbit
\beq
u=u^t \left(\partial_t+\frac{dr_p}{dt}\partial_r+\frac{d\phi_p}{dt}\partial_\phi\right)\,,
\eeq
where the expression ($\tau$ being the proper time parameter along the orbit)
\beq
u^t=\frac{dt_p}{d\tau}>0\,,
\eeq
(in order for $u$ to be future oriented) follows from the normalization condition $u\cdot u=-1$,
\beq
u^t=\frac{1}{\sqrt{f_s(r_p(t)){-}\frac{\left(\frac{dr_p}{dt}\right)^2}{f_s(r_p(t))f_b(r_p(t))}{-}r_p(t)^2  \left(\frac{d\phi_p}{dt}\right)^2}}\,.\qquad
\eeq
It is convenient to introduce   dimensionless   variables, e.g. $T$ defined as
\beq
T=\frac{vt}{b} \,,
\eeq
so that
\bea
r_p(T)&=&b\sqrt{1+T^2}\,,\nonumber\\ 
\phi_p(T)&=&{\rm arctan}\left(T\right)\,, 
\eea
which can also be easily inverted.

Expanding in powers of the dimensionless and PM-small quantity
\beq
\epsilon = GM/bv^2  
\eeq
(recall $M_{\rm TS} = \frac{M}{2}(\alpha+2)$) and the bookkeeping $\eta \sim 1/c$ (i.e., replacing $v$ with $\eta v$) one finds
\bea
u^t(T)&=&\gamma+\epsilon u^t_1+\epsilon^2 u^t_2+O(\eta^6,\epsilon^3)\,,
\eea
with $\gamma=1+\frac12 v^2\eta^2 +\frac38 v^4\eta^4+O(\eta^5)$ and
\bea
u^t_1&=&\frac{v^2\eta^2}{\sqrt{1+T^2}}+v^4\eta^4 \left(-\frac{1+\alpha}{(1+T^2)^{3/2}} +\frac{5+2\alpha}{2\sqrt{1+T^2}}\right)\,, \nonumber\\ 
u^t_2&=& \frac32 \frac{v^4\eta^4}{1+T^2}\,.
\eea
where $\alpha = r_b/r_s$ (introduced before).

As anticipated, the orbit turns out to be accelerated, with acceleration 
\beq
a(u)=\nabla_u u=a(u)^\alpha \partial_\alpha \,,
\eeq
with nonvanishing components ($\epsilon$ and $\eta$-expanded up to $O(\epsilon^2,\eta^5)$ for simplicity)
\bea
a(u)^t&=& \frac{1}{b}\frac{2 T v^3\epsilon\eta^3}{(T^2 + 1)^{3/2}}\,,\nonumber\\
a(u)^r&=& \frac{1}{b}\left[\frac{\epsilon v^2\eta^2}{(T^2 + 1)}+\left(-\frac{\epsilon  \alpha}{(T^2 + 1) }\right.\right.\nonumber\\
&-&\left. \left. 
\frac{2\alpha \epsilon^2}{(T^2 + 1)^{3/2} } 
- \frac{-3(1+\alpha)\epsilon}{ (T^2 + 1)^2)}\right)v^4\eta^4\right]\,.\qquad
\eea

Let us consider the main integral \eqref{main_integ} evaluating separately the building blocks
$A^m_1(t,t')$ and $A^{l\omega}_2(t,t')$ in a leading order (LO) approximation.
At the leading PN order (with $\gamma=1+O(\eta^2)$) we find
\beq
A^l_2(t,t')=-\frac{r_p(t)^{l}r_p(t')^{-l-1}}{(2 l+1)}\,,
\eeq
i.e., $A^{l\omega}_2(t,t')$ does not depend on $\omega$. Similarly, at LO $A^m_1(t,t)$ is independent of $m$ (the exponential factor cancels for $t'=t$). Therefore one has
\bea
\label{main_integ 3}
{\mathcal A}_{l}(t,t')|_{\rm LO}
&=&A_1(t,t) A^l_2(t,t)\delta(t'-t)\,,
\eea
with
\bea
A_1(t,t)&=&\frac{\sqrt{f_b(r_p(t)} }{u^t(t)}=1+O(\eta^2)\,,\nonumber\\
A^l_2(t,t)&=&-\frac{r_p(t)^{-1}}{(2 l+1)}+O(\eta^2)\,.
\eea
Consequently,
\bea
\label{main_integ 4}
{\mathcal A}_{l}(t,t')|_{\rm LO}
&=&-\frac{1}{\gamma (2 l+1)\sqrt{b^2+v^2t^2}}\delta(t'-t) \,,\qquad
\eea
and then, recalling that $\gamma=1+O(\eta^2)$ and
\beq
\sum_{m=-l}^l|Y_{lm}(\frac{\pi}{2},0)|^2=\frac{2l+1}{4\pi}\,,
\eeq
we find
\bea
\label{psi_val_TS_orbit_LO}
\psi|_{\rm orb, LO}
&=&4\pi q \sum_{lm} \frac{|Y_{lm}(\frac{\pi}{2},0)|^2}{(2 l+1)\sqrt{b^2+v^2t^2}}\nonumber\\
&=& q  \sum_{l=0}^\infty \frac{1}{\sqrt{b^2+v^2t^2}}\nonumber\\
&=& \sum_{l=0}^\infty \psi_l\,.
\eea
This sum clearly diverges since the terms $\psi_l$ actually are independent of  $l$. Therefore, the LO order term just computed will disappear when subtracting the singular field, according to a well established procedure (see e.g., \cite{Barack:2018yvs}).

It is instructive, however, to evaluate its FT, which turns out to be given by
\bea
\hat \psi_l(\omega) &=&  q \int dt \frac{e^{i\omega t}}{\sqrt{b^2+v^2t^2}}\nonumber\\
&=& \frac{2 q}{ v} K_0 (\hat \omega)\,,
\eea
where 
\beq
\hat \omega= \frac{\omega b}{v}\,.
\eeq
and $K_0$ denotes the Bessel K function of order 0.
This simple LO study is then implemented within the GSF formalism.
Using the PN solution and the $l=0$ only MST type solution the final results for the subtraction terms, usually called $B$-terms, read
\bea
B&=&\frac{q}{b}\left\{\frac{1}{\sqrt{T^2+1}}-\frac{\eta ^2 v^2}{4\left(T^2+1\right)^{3/2}}\right.\nonumber\\ 
&-&\eta ^4v^4
   \left(\frac{ \epsilon }{2 \left(T^2+1\right)^2}+\frac{1}{4 \left(T^2+1\right)^{3/2}}\right.\nonumber\\
&-&\left. \frac{9}{64 \left(T^2+1\right)^{5/2}}\right)\nonumber\\
&+&\eta ^6v^6 \left[\epsilon  \left(\frac{(8 \alpha +17)}{16 \left(T^2+1\right)^3}-\frac{(\alpha +3)}{2 \left(T^2+1\right)^2}\right)\right.\nonumber\\
&-&\frac{\epsilon^2}{\left(T^2+1\right)^{5/2}}-\frac{1}{4 \left(T^2+1\right)^{3/2}}+\frac{9}{32 \left(T^2+1\right)^{5/2}}\nonumber\\
&-&\left.\left.\frac{25}{256 \left(T^2+1\right)^{7/2}}\right]
\right\}+O(\eta^8)\,,
\eea
and in the Fourier domain
\bea
\hat B(\hat \omega)&=&\frac{q}{b}\left\{
-\frac{1}{2} \eta ^2 v^2 \hat \omega  K_1(\hat \omega )+2 K_0(\hat \omega ) \right.\nonumber\\
&+& \eta ^4 v^4\left(\frac{\hat \omega}{32}   (3\hat \omega K_0(\hat \omega )-10 K_1(\hat \omega ))\right.\nonumber\\
&-&\left. \frac{\pi e^{-\hat \omega }\epsilon }{4}    (\hat \omega  +1) \right)\nonumber\\
&+&
\eta ^6 v^6  \left(\frac{\pi e^{-\hat \omega } \epsilon}{128}   (8 \alpha  ((\hat \omega  -1) \hat \omega  -1)\right.\nonumber\\
&+&\hat \omega   (17 \hat \omega  -45)-45)\nonumber\\
&-& \frac{1}{384}  \hat \omega \left(\left(5 \hat \omega 
   ^2+88\right) K_1(\hat \omega )-52 \hat \omega  K_0(\hat \omega  )\right)\nonumber\\
&-&\left.\left.\frac{2}{3}  \hat \omega ^2 \epsilon ^2 K_2(\hat \omega )\right)
\right\}+O(\eta^8)\,.
\eea

The time-domain expression for the reconstructed field is given by
\bea
\psi(T)&=&\frac{q}{b}\epsilon  (\alpha +2)\left[
\eta ^4v^4 \left(\frac{3}{2 \left(T^2+1\right)^2}-\frac{1}{T^2+1}\right)\right.\nonumber\\
&+&\eta ^6 v^6   \left(
-\frac{5 }{6 \left(T^2+1\right)}
+\frac{35 }{12\left(T^2+1\right)^2}\right.\nonumber\\
&-&\left.\left. \frac{25 }{12 \left(T^2+1\right)^3}
\right)\right]+O(\epsilon^2,\eta^6)\,,
\eea
(here and below the symbol $O(\epsilon^2,\eta^6)$ denotes the corresponding terms at order $\epsilon^2$ computed including all $O(\eta^6)$ terms) 
and its frequency domain reconstruction
\bea
\hat \psi(\hat \omega)&=&\frac{q}{b}\pi  (\alpha+2)\epsilon e^{-\hat \omega} \left[ \frac{1}{4}  \eta ^4 v^4 (3 \hat \omega -1) \right.\nonumber\\
&+&\left. \eta ^6v^6   \frac{5}{96}   [\hat \omega(5 \hat \omega -13)+3] \right]+O(\epsilon^2,\eta^6)\,.\qquad
\eea
In the (soft) limit $\hat \omega \to 0$ the above relation reduces to
\bea 
\hat \psi(\hat \omega)&=& -\frac{q}{b}\pi  (\alpha +2)\epsilon \left[
\frac{1}{4} \eta ^4 v^4 (1-4\hat \omega) \right.\nonumber\\
&-&\left. \eta ^6 v^6 \frac{5}{96}(3-16\hat \omega)  \right]+ O(\hat \omega^2)\,.
\eea
Note that up to $O(\eta^6)$
\bea 
\hat \psi(0)&=& -\frac{q}{b}\pi  (\alpha +2)\epsilon \left[\frac{1}{4} \eta ^4 v^4  -\eta ^6 v^6 \frac{5}{32}   \right]\nonumber\\
&=& \hat \psi^{\epsilon^1,\eta^4}(0) \eta^4 +\hat \psi^{\epsilon^1,\eta^6}(0) \eta^6\,,
\eea
so that, for instance,
\beq
\frac{\hat \psi^{\epsilon^1,\eta^6}(0)}{\hat \psi^{\epsilon^1,\eta^4}(0)}=-\frac{5}{8}v^2 \,.
\eeq

\subsection{Equatorial geodesic unbound motion}

Geodesics in a TS spacetimes have largely studied.
In the equatorial case and for unbound orbits the radial equation reads \cite{Bianchi:2024vmi}
\bea
\label{eq_dr_dt}
\frac{dr}{dt}&=&\frac{\sqrt{r-r_b}}{\hat E  r^2}f_s(r)\sqrt{r^3(\hat E^2-1)+r_s r^2 -\hat L^2(r-r_s)}\,,\nonumber\\
\eea
with $\hat E=E/\mu$ dimensionless and the angular momentum $\hat L=J/\mu$  having the dimensions of a length ($j=J/(GM \mu)$ dimensionless).
When $r_b\to 0$ and $r_s\to 0$ Eq. \eqref{eq_dr_dt} reduces to the flat space equation
\bea
\frac{dr}{dt}\big|_{\rm flat}&=&\frac{\sqrt{r}}{\hat Er^2}\sqrt{r^3(\hat E^2-1)-\hat L^2r}\nonumber\\
&=&\frac{1}{\hat E}\sqrt{\hat E^2-1-\frac{\hat L^2}{r^2}}\,.
\eea
Using the notation\footnote{Here $\gamma$ is the Lorentz factor and should not be confused with the Euler-Mascheroni constant also denoted by the same symbol.} 
\beq
\label{notation_gamma_b}
\hat E=\gamma=(1-v^2)^{-1/2}\,,\qquad \hat L=b \sqrt{\hat E^2-1}=b \gamma v\,,
\eeq
one finds
\bea
\frac{dr}{dt}\bigg|_{\rm flat} &=&\frac{1}{\gamma}\sqrt{\gamma^2v^2-b^2\frac{\gamma^2v^2}{r^2}}\nonumber\\
&=& \frac{v}{r}\sqrt{r^2-b^2}\,,
\eea
which integrates to
\beq
r\big|_{\rm flat}=\sqrt{b^2+v^2 t^2}\,.
\eeq

Let us rewrite Eq. \eqref{eq_dr_dt} as follows
\bea
\label{eq_dr_dt2}
\frac{dr}{dt} 
&=& \frac{\sqrt{f_b(r)}}{\hat E}f_s(r)\sqrt{ \hat E^2-f_s(r) \left(1+ \frac{\hat L^2}{r^2}\right)}\,.
\eea
Using Eq. \eqref{notation_gamma_b} we obtain the equivalent expression
\bea
\label{eq_dr_dt3}
\frac{dr}{dt}
&=& v \sqrt{f_b(r)}f_s(r)\sqrt{1-\frac{b^2}{r^2} +\frac{r_s}{r}\left(\frac{1}{\gamma^2 v^2}   +\frac{b^2}{r^2}\right) }\,,\nonumber\\
\eea
which, indeed, manifestly reduces to the flat space limit 
\bea
\label{eq_dr_dt3_flat}
\frac{dr}{dt}\bigg|_{\rm flat}&=& v\sqrt{1-\frac{b^2}{r^2} }\,,
\eea
when $r_s\to 0$ and $r_b\to 0$.

The remaining  geodesic equations written by using the coordinate time as a parameter are given by
\bea
\frac{d\phi}{dt}&=&\frac{\hat L}{\hat E r^2}\left(1-\frac{r_s}{r} \right)\,,\nonumber\\
\frac{dt}{d\tau}&=& \frac{\hat E}{1-\frac{r_s}{r} }\,,
\eea
and can be easily integrated once the solution $r=r(t)$ is known. Alternately, one can use the `radial' action $S_r(r, r_0;E,J)= \int_{r_0}^r p_r(r;E,J) dr$ and get 
\beq
\phi=\phi_0 + {\partial S_r\over \partial J}\,, \quad t=t_0 - {\partial S_r\over \partial E} \,.
\eeq

Using the (already defined) dimensionless time parameter $T={vt}/{b}$
and the dimensionless PM expansion parameter $\epsilon$ ($\epsilon\ll 1$)
\beq
\epsilon=\frac{GM}{bv^2}=\frac{Gr_s}{2bv^2}\,, 
\eeq
i.e.,  $\epsilon={r_s}/{2bv^2}$ when using $G=1$,
we can write the solutions of the geodesic equations in the following PM-expanded form
\bea
\label{PM_sol}
r(T)&=& b \sqrt{1+T^2}+\epsilon r_1+\epsilon^2 r_2+O(\epsilon^3)\,,\nonumber\\
\phi(T)&=& {\rm arctan}(T) +\epsilon \phi_1+\epsilon^2 \phi_2+O(\epsilon^3) \,,\nonumber\\
\frac{dt}{d\tau} &=& \gamma  +\epsilon u^t_1 +\epsilon^2 u^t_2+O(\epsilon^3)\,,
\eea
with $\gamma=1+\frac{\eta ^2 v^2}{2}+\frac{3 \eta ^4 v^4}{8}+O(\eta^6)$ and
where the various coefficients, also having a PN expansion, are listed in Table \ref{tab:ri-phii-uti} below.

\begin{table*}  
\caption{\label{tab:ri-phii-uti} List of the various coefficients entering the PM expansion of the unbound geodesics in the TS spacetime, Eq. \eqref{PM_sol}, in powers of $\epsilon=\frac{r_s}{2bv^2}$.
Note that $r(0)=r_{\rm min}$, i.e., $T=0$ corresponds to the turning point, where the minimum approach distance is reached by the massive probe,
$r_{\rm min}=b \left[1-\epsilon +\frac12 \epsilon^2 (1-4v^2\eta^2) \right]+O(\epsilon^3)$.
}
\begin{ruledtabular}
\begin{tabular}{ll}
$\frac{r_1}{b}$ & $ -1+\frac{T {\rm arcsinh}(T)}{\sqrt{1+T^2}}-v^2\eta^2\frac{(\alpha +3)   T {\rm arcsinh}(T)}{\sqrt{1+T^2}} $\\
$\frac{r_2}{b}$ & $\frac{1}{2 \sqrt{T^2+1}}+\frac{T {\rm arcsinh}(T)}{1+T^2}+\frac{{\rm arcsinh}(T)^2}{2
   \left(1+T^2\right)^{3/2}}+v^2\eta^2\Big[-\frac{2}{\sqrt{1+T^2}}-\frac{2
   (\alpha +3) T {\rm arcsinh}(T)}{1+T^2}-\frac{(\alpha +3) {\rm arcsinh}(T)^2}{\left(1+T^2\right)^{3/2}}\Big]$\\
   $$ & $+v^4 \eta^4\Big[ -\frac{3 \left(\alpha ^2+2 \alpha +5\right) T {\rm arctan}(T)}{2
   \sqrt{1+T^2}}+\frac{(\alpha +3)^2 T {\rm arcsinh}(T)}{1+T^2}+\frac{(\alpha +3)^2 {\rm arcsinh}(T)^2}{2
   \left(1+T^2\right)^{3/2}}\Big]$\\
 \hline
$\phi_1$ & $\frac{T}{\sqrt{1+T^2}}+\frac{{\rm arcsinh}(T)}{1+T^2}+v^2 \eta^2\Big[\frac{(\alpha +1) T }{\sqrt{1+T^2}}-\frac{(\alpha +3)  {\rm arcsinh}(T)}{1+T^2}\Big]$\\
$\phi_2$ & $\frac{2 {\rm arcsinh}(T)}{\left(1+T^2\right)^{3/2}}-\frac{T {\rm arcsinh}(T)^2}{\left(1+T^2\right)^2}+v^2 \eta^2\Big[\frac{(\alpha +3) T}{1+T^2}+\frac{2 (\alpha +3) T {\rm arcsinh}(T)^2}{\left(1+T^2\right)^2}-\frac{2 (\alpha +4) {\rm arcsinh}(T)}{\left(1+T^2\right)^{3/2}}+(\alpha +3) {\rm arctan}(T)\Big] $\\
&$v^4 \eta^4\Big[\frac{\left(3 \alpha ^2+2 \alpha +3\right) T}{4 \left(1+T^2\right)}-\frac{3
   \left(\alpha ^2+2 \alpha +5\right) {\rm arctan}(T)}{2
   \left(1+T^2\right)}-\frac{(\alpha +3)^2 T {\rm arcsinh}(T)^2}{\left(1+T^2\right)^2}+\frac{2 (\alpha +3) {\rm arcsinh}(T)}{\left(1+T^2\right)^{3/2}}+\frac{1}{4} \left(3 \alpha ^2+2
   \alpha +3\right) {\rm arctan}(T)\Big]$\\
\hline
$u^t_1$ &$-\frac{(\alpha -2) \eta ^2
   v^2}{\sqrt{1+T^2}}-\frac{(\alpha -2) \eta ^4 v^4}{2 \sqrt{1+T^2}}$\\
$u^t_2$ & $\eta ^2 v^2\left(-\frac{(\alpha -2) }{1+T^2}+\frac{(\alpha -2) T  {\rm arcsinh}(T)}{\left(1+T^2\right)^{3/2}}\right)+v^4\eta^4\Big[ -\frac{\alpha ^2+5 \alpha -10 }{2
   \left(1+T^2\right)}-\frac{\left(2 \alpha ^2+\alpha -10\right) T  {\rm arcsinh}(T)}{2 \left(1+T^2\right)^{3/2}}\Big]$\\
\end{tabular}
\end{ruledtabular}
\end{table*}

\begin{widetext}
We can now repeat the previous computation, Eq. \eqref{psi_val_TS},  namely proceed with the following tasks: 1) evaluate the scalar field at a generic spacetime point
\bea
\label{psi_val_TS_1}
\psi(t,r,\theta,\phi)
&=&-4\pi q \sum_{lm}Y_{lm}(\theta,\phi)Y_{lm}^*(\frac{\pi}{2},0)\int dt'  \int \frac{d\omega}{2\pi}e^{-i\omega t}   G_{l{}\omega}(r,r_p(t'))
 \frac{e^{i(\omega t'-m\phi_p(t'))} \sqrt{f_b(r_p(t'))} }{u^t(t')} \,.
\eea
\end{widetext}
Here, in principle, one should compute the left and right fields (approaching the particle source world line from left and from right) since we have either a left or a right Green function.
However, due to  the continuity of the field $\psi^+ = \psi^-$ along the particle's world line, we can use the left (or the right) field. 
2) specialize the result of 1)  along the (geodesic) hyperbolic-like orbit, i.e. to evaluate  $\psi_{\rm orb}(t)=\psi(t,r_p(t),\frac{\pi}{2},\phi_p(t))$
\bea
\label{psi_val_TS_orbit_1}
\psi_{\rm orb}(t)
&=&-4\pi q \sum_{lm}|Y_{lm}(\frac{\pi}{2},0)|^2  \int dt'  {\mathcal A}_{lm}(t,t')\,,\qquad
\eea
with
\bea
\label{main_integ_1}
{\mathcal A}_{lm}(t,t')
&=& A^m_1(t,t') \int \frac{d\omega}{2\pi} e^{-i\omega (t-t')}A^{l\omega}_2(t,t') \,,\qquad 
\eea
and, following our previous notation,
\bea
A^m_1(t,t')&=& e^{i m (\phi_p(t)-\phi_p(t'))}\frac{\sqrt{f_b(r_p(t')} }{u^t(t')}\nonumber\\
A^{l\omega}_2(t,t')&=& \frac{R_{\rm in}^{l{}\omega}(r_p(t))R_{\rm up}^{l{}\omega}(r_p(t'))}{W_{l{}\omega}}\,.
\eea
Noticeably, $A^{l\omega}_2(t,t')$ when computed with the PN solutions for $R^{l\omega}_{\rm in/up}$ does not depend on $m$ and it is an even polynomial in $\omega$ of the type
\bea
A^{l\omega, {\rm PN}}_2(t,t')&=& A^{l\omega^0}_2(t,t')+\omega^2 A^{l\omega^2}_2(t,t')+\omega^4 A^{l\omega^4}_2(t,t')\nonumber\\
&+& O(\omega^6)\,.
\eea
Notation here is the following: $ A^{l\omega^n}_2$ is the coefficient of $\omega^n$ in an expansion in powers of $\omega$ but it does not depend on $\omega$ anymore.  
Differently, when computed with the MST solutions for $R^{l{}\omega}_{\rm in/up}$, $ A^{l}_{2 \, {\rm MST} }(t,t')$ does not depend on $m$ either but has an expansion in $\omega$ which involves log's and powers of $\omega$, e.g. for $l=0$
\bea
A^{l=0}_{2\,\rm MST}(t,t')&=& A_2^{l=0,\omega^0}(t,t')+\omega A_2^{l=0,\omega^1}(t,t')\nonumber\\
&+&\omega^2 [A_2^{l=0,c,\omega^2}(t,t')+A_2^{l=0,\ln{},\omega^2}(t,t')\ln \omega]\nonumber\\
&+& O(\omega^3,\ln \omega)\,.
\eea

Finally, 3)  compute the Fourier transform of the field along the orbit.

Moreover,  the solutions (PN, MST) of the radial equation for the scalar field motion in the TS spacetime exist only in PN sense: we need then to take the PN limit of our PM results. Restoring physical units, i.e.  
replacing $M\to M\eta^2$ (and similarly $M_{\rm TS} \to M_{\rm TS} \eta^2$ since both $r_s$ and $r_b$ rescale in the same way for $G=1$), $v\to v \eta$, one expands in $\eta$ up to the desired PN accuracy, besides PM expanding all expressions where PM evaluated quantities enter.

Therefore
\begin{widetext}
\bea
\int \frac{d\omega}{2\pi} e^{-i\omega (t-t')}A_2^{l\omega \rm PN}(t,t')&=& \int \frac{d\omega}{2\pi} e^{-i\omega (t-t')}\left[A_2^{l \omega^0}(t,t')+\omega^2 A_2^{l \omega^2}(t,t')+\omega^4 A_2^{l \omega^4}(t,t')+O(\omega^6)\right]\nonumber\\
&=& A_2^{l \omega^0}(t,t') \delta(t-t')+A_2^{l \omega^2}(t,t')[-\frac{d^2}{dt^2}\delta(t-t')]+A_2^{l \omega^4}(t,t')[\frac{d^4}{dt^4}\delta(t-t')]+\ldots \nonumber\\
&=& \delta(t-t')\widehat A^l_2(t)\,,
\eea
with $\widehat A^l_2(t)$ (depending on $l$) given by
\beq
\widehat A^l_2(t)=A_2^{l\omega^0}(t,t) -\frac{d^2}{dt^2}A_2^{l \omega^2}(t,t')\bigg|_{t'=t}  +\frac{d^4}{dt^4}A_2^{l \omega^4}(t,t')\bigg|_{t'=t} +\ldots \,.
\eeq
\end{widetext}
This implies
\bea
\label{main_integ_1comput}
{\mathcal A}_{lm}(t,t')
&=& A_1(t,t) \widehat A^l_2(t)\delta(t-t')  \,,\qquad 
\eea
independent of $m$ and finally
\bea
\label{psi_val_TS_orbit_1comput}
\psi_{\rm orb}(t)
&=&-4\pi q \sum_{l}\frac{2l+1}{4\pi}  A_1(t,t) \widehat A_2(t) \nonumber\\
&=&-  q \frac{\sqrt{f_b(r_p(t))} }{u^t(t)} \sum_{l} (2l+1) \widehat A^l_2(t)\,,\qquad
\eea
since also
\beq
A_1(t,t)=\frac{\sqrt{f_b(r_p(t))} }{u^t(t)}\,,
\eeq
does not depend on $m$.
We notice the following two features: 1) the sum
$\sum_{l} (2l+1) \widehat A^l_2(t)$ diverges and, according to a standard procedure already mentioned,  one has to regularize it by subtracting the contribution of the singular field. This implies the identification of subtraction terms, called $B$-terms, fully obtained by using the PN solution. 
When using the MST solutions, instead, one has expressions involving either \lq\lq constant" ($c-$type terms) PN-type terms 
\beq
\sum_{k}A_2^{l=n,{\rm c},\omega^k}(t,t')\int \frac{d\omega}{2\pi} e^{-i\omega (t-t')}\omega^k\,,
\eeq
and new terms involving $\ln \omega$  and powers of $\omega$, ($\ln$-type terms),
\beq
A_2^{l=n,\ln{},\omega^k}(t,t')\int \frac{d\omega}{2\pi} e^{-i\omega (t-t')}\omega^k \ln \omega\,,
\eeq
that is
\begin{widetext}
\bea
\int \frac{d\omega}{2\pi} e^{-i\omega (t-t')}A_2^{l=n,\rm MST}(t,t')&=& \sum_{k}A_2^{l=n,{\rm c},\omega^k}(t,t')\int \frac{d\omega}{2\pi} e^{-i\omega (t-t')}\omega^k\nonumber\\
&+&A_2^{l=n,\ln{},\omega^2}(t,t')\int \frac{d\omega}{2\pi} e^{-i\omega (t-t')}\omega^2 \ln \omega\nonumber\\
&+& A_2^{l=n,\ln{},\omega^3}(t,t')\int \frac{d\omega}{2\pi} e^{-i\omega (t-t')}\omega^3 \ln \omega+\ldots\nonumber\\
&=& \sum_{k}A_2^{l=n,{\rm c},\omega^k}(t,t') \frac{1}{(-i)^k } \delta^{(k)}(t-t')\nonumber\\ 
&-&A_2^{l=n,\ln{},\omega^2}(t,t')\frac{d^2}{dt^2}W(t-t')+i A_2^{l=n,\ln{},\omega^3}(t,t')\frac{d^3}{dt^3}W(t-t')+\ldots
\eea
where
\beq
W(t-t')=\int \frac{d\omega}{2\pi} e^{-i\omega (t-t')}\ln \omega\,.
\eeq
The function $W(x)$ can be evaluated to be \cite{Racine:2008kj}  
\bea
\label{Wx}
W(x)&=&\int \frac{d\omega}{2\pi} e^{-i\omega x}\ln \omega\nonumber\\
&=& -\frac{1}{x}H(x)+\left(\frac{i\pi}{2}- \gamma \right) \delta(x)\,.
\eea
Appendix \ref{appB} contains details on the evaluation of this and similar integrals.
We then find 
\bea
\int _{-\infty}^{\infty} dt' A(t,t') \int_{-\infty}^\infty \frac{d\omega}{2\pi}  e^{-i\omega (t-t')}\ln \omega&=& 
-\int_{-\infty}^{t} dt' \frac{A(t, t')}{t-t'}+\left(\frac{i\pi}{2}- \gamma \right) \int_{-\infty}^{\infty} dt' A(t,t')\delta(t-t')\nonumber\\
&=& -\int_{-\infty}^{t} dt' \frac{A(t, t')}{t-t'}+\left(\frac{i\pi}{2}- \gamma \right)  A(t,t)\nonumber\\
&\equiv & C(t)+\left(\frac{i\pi}{2}- \gamma \right)  A(t,t)\,.
\eea
We see that the non-delta part of $W(t-t')$ introduces further complications, in the sense that one has to perform integrals of the type
\beq
C(t)=-\int_{-\infty}^{t} dt' \frac{A(t, t')}{t-t'}\,,
\eeq
which, in general, are singular at $t=t'$, and then the evaluation of the associated finite part is necessary too. A first, direct route is to introduce a (dimensionless) scale $\sigma>0$, and evaluate $C(t)$ as $C_\sigma(t)$ according to a cut-off regularization,
\bea
C_\sigma(t)&=&-\int_{-\infty}^{t} dt' \frac{A(t, t')}{t-t'}\nonumber\\
&=&-\frac{b}{v}\int_{-\infty}^{T} dT' \frac{A(t=\frac{bT}{v}, t'=\frac{bT'}{v})}{\frac{b}{v}(T-T')}\bigg|_{T=\frac{vt}{b}}\nonumber\\
&=& - \int_{-\frac{1}{\sigma}}^{T-\sigma} dT' \frac{A(t=\frac{bT}{v}, t'=\frac{bT'}{v})}{(T-T')}\bigg|_{T=\frac{vt}{b}}\,,
\eea
with $\sigma$ dimensionless. Taking the finite part of a divergent integral is equivalent to evaluate the series in small $\sigma$, keeping only finite contributions as well as all log-divergent terms which will be used to introduce a scale in the intermediate results (which however will cancel in the evaluation of any gauge-invariant quantity).
For example, 
\bea
C_\sigma(t)&=&\ldots \frac{{{C}}_{-3}}{\sigma^3}+\frac{{{C}}_{-2}}{\sigma^2}+\frac{{{C}}_{-1}}{\sigma}+{{C}}_0+ {{C}}_{\ln}\ln \sigma +{{C}}_1 \sigma +{{C}}_2\sigma^2+\ldots\nonumber\\
&=& {{C}}_0+ {{C}}_{\ln}\ln \sigma+\ldots\,,
\eea
including higher powers of logs when appearing.
The final result associated with a divergent integral will depend anyway on $\sigma$.
A second route, more solid from a mathematical point of view,  consists in evaluating $C_\sigma(t)$ as the \lq\lq Hadamard Partie Finie" of the corresponding integral,
\bea
C_\sigma (t)&=&-\int_{-\infty}^{t} dt' \frac{A(t, t')}{t-t'}\nonumber\\
&=& -\int_{-\infty}^{t} dt' \frac{A(t, t')-A(t,t)}{t-t'}-A(t,t){\rm PF}_\sigma \int_{-\infty}^{t} dt' \frac{1}{t-t'} \nonumber\\
&=& -\int_{-\infty}^{t} dt' \frac{A(t, t')-A(t,t)}{t-t'}-A(t,t) {\rm PF}_\sigma \int_0^\infty \frac{d\xi}{\xi}\,.
\eea  
Here, the first integral, in general, is no more divergent at $t'=t$, whereas the last integral diverges both in $0$ and at infinity, and can be performed from $\sigma$ up to $1/\sigma$ instead of $0$ and $\infty$, leading to
\beq
{\rm PF}_\sigma \int_0^\infty \frac{d\xi}{\xi}=-2\ln \sigma\,.
\eeq
We include in Appendix \ref{PF_recalls} a short discussion on this type of integrals.
We end this section by writing explicitly the $B-$term
\bea
B &=& \sum_{n=0}^\infty B^{\epsilon^n}(T)\,.
\eea
In order to have a compact expression of the $B-$term
it can be convenient to introduce the following notation
\bea
f_n &=& \frac{1}{(T^2+1)^n}\,,\qquad \tilde f_n=\frac{T}{(T^2+1)^n}\,\qquad f_n^{\rm ask}= \frac{{\rm arcsinh}^k(T)}{(T^2+1)^n}\,, \nonumber\\
\tilde{f}_n^{\rm ask}&=& \frac{T{\rm arcsinh}^k(T)}{(T^2+1)^n} \,, \qquad f_n^{\rm at1}= \frac{{\rm arctan}(T)}{(T^2+1)^n}\,,\qquad \tilde{f}_n^{\rm at1}= \frac{T{\rm arctan}(T)}{(T^2+1)^n}\,,\nonumber\\
&&\qquad \qquad \qquad \qquad \quad f_n^{\rm log}=\frac{{\rm Log}(1+T^2)}{2(1+T^2)^n}\,.
\eea
Limiting our computations to $O(\eta^6)$ included,  we find 
\bea
B^{\epsilon^0}(T)&=& \frac{q}{b}\Bigg[f_{\frac{1}{2}}-\frac{1}{4} f_{\frac{3}{2}} \eta ^2 v^2+\eta ^4 v^4 \left(\frac{9}{64} f_{\frac{5}{2}} -\frac{1}{4} f_{\frac{3}{2}}
   \right)+\eta ^6 v^6 \left(-\frac{1}{4} f_{\frac{3}{2}} +\frac{9}{32} f_{\frac{5}{2}}
   -\frac{25}{256} f_{\frac{7}{2}}\right)\Bigg]\,,\nonumber\\
B^{\epsilon^1}(T)&=& \frac{q}{b}\Bigg[f_1+\tilde{f}_{\frac{3}{2}}^{\text{as1}}+\eta ^2 v^2 \left(-\alpha  \tilde{f}_{\frac{3}{2}}^{\text{as1}}-3
   \tilde{f}_{\frac{3}{2}}^{\text{as1}}-\frac{3}{4}
   \tilde{f}_{\frac{5}{2}}^{\text{as1}}-\frac{3 f_2}{4}\right)+\eta ^4 v^4 \left(\frac{3}{4} \alpha 
   \tilde{f}_{\frac{5}{2}}^{\text{as1}}+\frac{3}{2}
   \tilde{f}_{\frac{5}{2}}^{\text{as1}}+\frac{45}{64}
   \tilde{f}_{\frac{7}{2}}^{\text{as1}}\right.\nonumber\\
   &-&\left.\frac{3 f_2}{4}{+}\frac{45 f_3}{64}\right){+}\eta ^6 v^6 \left(\frac{3}{4} \alpha 
   \tilde{f}_{\frac{5}{2}}^{\text{as1}}{-}\frac{45}{64} \alpha 
   \tilde{f}_{\frac{7}{2}}^{\text{as1}}{+}\frac{3}{2}
   \tilde{f}_{\frac{5}{2}}^{\text{as1}}{-}\frac{45}{64}
   \tilde{f}_{\frac{7}{2}}^{\text{as1}}{-}\frac{175}{256}
   \tilde{f}_{\frac{9}{2}}^{\text{as1}}{-}\frac{3 f_2}{4}{+}\frac{45 f_3}{32}{-}\frac{175
   f_4}{256}\right)\Bigg]\,.
\eea
Similarly, at $O(\epsilon^2)$  we find
\bea\label{btermepsilon2}
B^{\epsilon^2}(T)&=& B^{\epsilon^2,\eta^0}(T)+\eta^2 v^2B^{\epsilon^2,\eta^2}(T)+\eta^4 v^4 B^{\epsilon^2,\eta^4}(T)+\eta^6 v^6 B^{\epsilon^2,\eta^6}(T)+O(\eta^8)\,,
\eea
where the coefficients appearing in \eqref{btermepsilon2} are collected in Table \ref{tabbeps2}.
\begin{table}[]
\caption{B-term coefficients as defined in \eqref{btermepsilon2}.}\label{tabbeps2}
\begin{tabular}{|c|l|}
\hline
$B^{\epsilon^2,\eta^0}(T)$                       & $\frac{q}{b}\left(3 \tilde{f}_2^{\text{as1}}-f_{\frac{3}{2}}^{\text{as2}}+\frac{3
   f_{\frac{5}{2}}^{\text{as2}}}{2}+\frac{f_{\frac{3}{2}}}{2}\right)$ \\ \hline
$B^{\epsilon^2,\eta^2}(T)$                       & $\frac{q}{v}\left(-4 \alpha  \tilde{f}_2^{\text{as1}}-12 \tilde{f}_2^{\text{as1}}-\frac{15
   \tilde{f}_3^{\text{as1}}}{4}+2 \alpha  f_{\frac{3}{2}}^{\text{as2}}-3 \alpha 
   f_{\frac{5}{2}}^{\text{as2}}+6 f_{\frac{3}{2}}^{\text{as2}}-\frac{15
   f_{\frac{5}{2}}^{\text{as2}}}{2}-\frac{15 f_{\frac{7}{2}}^{\text{as2}}}{8}+2
   f_{\frac{3}{2}}-\frac{9 f_{\frac{5}{2}}}{8}\right)$ \\ \hline
$B^{\epsilon^2,\eta^4}(T)$                       & $\frac{q}{b}\left(\alpha ^2 \tilde{f}_2^{\text{as1}}+6 \alpha  \tilde{f}_2^{\text{as1}}+\frac{9}{2}
   \alpha  \tilde{f}_3^{\text{as1}}+9 \tilde{f}_2^{\text{as1}}+\frac{39
   \tilde{f}_3^{\text{as1}}}{4}+\frac{315 \tilde{f}_4^{\text{as1}}}{64}-\frac{3}{2}
   \alpha ^2 \tilde{f}_{\frac{3}{2}}^{\text{at1}}-3 \alpha 
   \tilde{f}_{\frac{3}{2}}^{\text{at1}}-\frac{15}{2}
   \tilde{f}_{\frac{3}{2}}^{\text{at1}}-\alpha ^2
   f_{\frac{3}{2}}^{\text{as2}}\right.$\\
   &$\left.+\frac{3}{2} \alpha ^2 f_{\frac{5}{2}}^{\text{as2}}-6
   \alpha  f_{\frac{3}{2}}^{\text{as2}}+6 \alpha 
   f_{\frac{5}{2}}^{\text{as2}}+\frac{15}{4} \alpha  f_{\frac{7}{2}}^{\text{as2}}-9
   f_{\frac{3}{2}}^{\text{as2}}+6 f_{\frac{5}{2}}^{\text{as2}}+\frac{465
   f_{\frac{7}{2}}^{\text{as2}}}{64}+\frac{315
   f_{\frac{9}{2}}^{\text{as2}}}{128}-\frac{21 f_{\frac{5}{2}}}{8}+\frac{225
   f_{\frac{7}{2}}}{128}\right)$ \\ \hline
\multicolumn{1}{|l|}{$B^{\epsilon^2,\eta^6}(T)$} & $\frac{q}{b}\left(-\frac{3}{4} \alpha ^2 \tilde{f}_3^{\text{as1}}-\frac{45}{8} \alpha 
   \tilde{f}_4^{\text{as1}}+3 \tilde{f}_3^{\text{as1}}-\frac{225
   \tilde{f}_4^{\text{as1}}}{32}-\frac{1575
   \tilde{f}_5^{\text{as1}}}{256}+\frac{9}{8} \alpha ^2
   \tilde{f}_{\frac{5}{2}}^{\text{at1}}+\frac{9}{4} \alpha 
   \tilde{f}_{\frac{5}{2}}^{\text{at1}}+\frac{45}{8}
   \tilde{f}_{\frac{5}{2}}^{\text{at1}}+\frac{3}{2} \alpha ^2
   f_{\frac{5}{2}}^{\text{as2}}\right.$\\
   &$\left.-\frac{15}{8} \alpha ^2
   f_{\frac{7}{2}}^{\text{as2}}+6 \alpha 
   f_{\frac{5}{2}}^{\text{as2}}-\frac{105}{32} \alpha 
   f_{\frac{7}{2}}^{\text{as2}}-\frac{315}{64} \alpha 
   f_{\frac{9}{2}}^{\text{as2}}+6 f_{\frac{5}{2}}^{\text{as2}}+\frac{15
   f_{\frac{7}{2}}^{\text{as2}}}{16}-\frac{455
   f_{\frac{9}{2}}^{\text{as2}}}{64}-\frac{1575
   f_{\frac{11}{2}}^{\text{as2}}}{512}-\frac{21 f_{\frac{5}{2}}}{8}+\frac{315
   f_{\frac{7}{2}}}{64}-\frac{1225 f_{\frac{9}{2}}}{512}\right)$ \\ \hline
\end{tabular}
\end{table}
Clearly, 1) the above rewriting with the various $f_n^{\rm X}$ is better illustrative of the structure of the various terms; 2) we refrain from displaying here additional PN terms (which we computed anyway). 

The Fourier transform of the $B-$term is given by
\bea
\hat B^{\epsilon^0}(\hat \omega)&=&\frac{q}{b}\left[2  K_0(\hat{\omega} )-\frac{\eta ^2  v^2 \hat{\omega}  }{2 }K_1(\hat{\omega})
+\eta ^4 v^4\left(\frac{3   \hat{\omega} ^2}{32 } K_0(\hat{\omega} )-\frac{5  
   \hat{\omega}  }{16 }K_1(\hat{\omega} )\right)\right.\nonumber\\
   &+&\left.\eta ^6 v^6\left(\frac{13   \hat{\omega} ^2 }{96 }K_0(\hat{\omega} )-\frac{ 
   \hat{\omega}  \left(5 \hat{\omega} ^2+88\right) }{384 }K_1(\hat{\omega} )\right)\right]\,,
\eea
and
\bea
\hat B^{\epsilon^1,\eta^0}( \hat{\omega})&=&\frac{\pi  q \hat{\omega}  K_0(\hat{\omega} )}{b}\,, \nonumber\\
\hat B^{\epsilon^1,\eta^2}( \hat{\omega})&=& \frac{q}{b}v^2\left[\frac{1}{4} \pi  (4 \alpha +11) e^{-\hat{\omega} }-\pi  (\alpha +3) \hat{\omega} 
   K_0(\hat{\omega} )-\frac{1}{4} \pi  \hat{\omega} ^2 K_1(\hat{\omega} )\right]\,,\nonumber\\
\hat B^{\epsilon^1,\eta^4}(\hat \omega)&=& \frac{q}{b}v^4 \left[-\frac{1}{128} \pi  e^{-\hat{\omega }} \left(16 \alpha  \left(3 \hat{\omega
   }+1\right)+129 \hat{\omega }+53\right)+\frac{1}{32} \pi  (8 \alpha +19)
   \hat{\omega }^2 K_1\left(\hat{\omega }\right)+\frac{3}{64} \pi 
   \hat{\omega }^3 K_0\left(\hat{\omega }\right)\right]\,,\nonumber\\
   \hat B^{\epsilon^1,\eta^6}(\hat \omega)&=& \frac{q}{b}v^6 \left[\frac{\pi  e^{-\hat{\omega }} \left(12 \alpha  \left(45 \hat{\omega
   }^2-117 \hat{\omega }-37\right)+1445 \hat{\omega }^2-3577 \hat{\omega
   }-1401\right)}{6144}-\frac{1}{192} \pi  (9 \alpha +14) \hat{\omega }^3
   K_0\left(\hat{\omega }\right)\right.\nonumber\\
   &+&\left.\frac{1}{768} \pi  \hat{\omega }^2
   \left(120 \alpha -5 \hat{\omega }^2+272\right) K_1\left(\hat{\omega
   }\right)\right]\,.
\eea
As said above, $B^{\epsilon^2,\eta^n}(\hat \omega)$ are collected in the related Supplemental Material.

\subsection{The reconstructed field along the source world line: time domain computations}

We can now display the reconstructed field in $\epsilon-\eta$ expanded form, i.e., one of the main original results of the present work. 
As before, we decompose it in powers of $\epsilon$ (dimensionless), with each coefficient having an expansion in $\eta$, namely
\beq
\psi(T)= \psi^{\epsilon^0}(T)+\epsilon \psi^{\epsilon^1}(T)+\epsilon^2 \psi^{\epsilon^2}(T)+\ldots\,,
\eeq
with
\bea
\psi^{\epsilon^n}(T)&=&\psi^{(\epsilon^n,\eta^0)}(T)+\eta^2 \psi^{(\epsilon^n,\eta^2)}(T)+\eta^3\psi^{(\epsilon^n,\eta^3)}(T)+\eta^4\psi^{(\epsilon^n,\eta^4)}(T)+\eta^5\psi^{(\epsilon^n,\eta^5)}(T) +\ldots \,,
\eea
where $\psi^{\epsilon^n}(T)$, $n=0,1,2$ are 
\bea
\psi^{\epsilon^0}(T)&=&0\,,\nonumber\\
\psi^{\epsilon^1}(T)&=&\frac{q}{b}\Bigg[-(\alpha +1) \eta ^3 v^3 \tilde{f}_{\frac{3}{2}}-\frac{1}{2} (\alpha +2) \left(2 f_1-3 f_2\right) \eta ^4 v^4-\frac{1}{30} \eta ^5 v^5 \left((45 \alpha +83) \tilde{f}_{\frac{3}{2}}-10 (9 \alpha
   +14) \tilde{f}_{\frac{5}{2}}\right)\nonumber\\
   &-&\frac{1}{4} (\alpha +2) \left(6 f_1-25 f_2+20 f_3\right) \eta^6 v^6\Bigg]\,,\nonumber\\
\psi^{\epsilon^2}(T)&=&\frac{q}{b}\Bigg[\eta^3 v^3(\alpha +1)  \left(2 f_{\frac{3}{2}}^{\text{as1}}-3
   \left(\tilde{f}_2+f_{\frac{5}{2}}^{\text{as1}}\right)\right)-\eta^4 v^4 (\alpha +2) \left(-2 \tilde{f}_2^{\text{as1}}+6 \tilde{f}_3^{\text{as1}}+5
   f_{\frac{3}{2}}-6 f_{\frac{5}{2}}\right)\nonumber\\
   &-&\eta^5 v^5\frac{1}{90} \left(495 \alpha  \tilde{f}_2-1350 \alpha  \tilde{f}_3+1817
   \tilde{f}_2-2100 \tilde{f}_3+6 \left(30 \alpha ^2+75 \alpha +7\right)
   f_{\frac{3}{2}}^{\text{as1}}\right.\nonumber\\
   &+&\left. \left(-270 \alpha ^2+405 \alpha +1617\right)
   f_{\frac{5}{2}}^{\text{as1}}-1350 \alpha  f_{\frac{7}{2}}^{\text{as1}}-2100
   f_{\frac{7}{2}}^{\text{as1}}\right)-\frac{\eta^6v^6}{288}\left(96 (\alpha +2) \left(6 \alpha  \tilde{f}_2^{\text{as1}}\right.\right.\nonumber\\
   &-&\left.\left. 18 \alpha 
   \tilde{f}_3^{\text{as1}}+9 \tilde{f}_2^{\text{as1}}+21
   \tilde{f}_3^{\text{as1}}-90 \tilde{f}_4^{\text{as1}}+18 \alpha  \tilde{f}_2+12
   \tilde{f}_2-4 (3 \alpha +2) f_{\frac{3}{2}}^{\text{as1}}+18 \alpha 
   f_{\frac{5}{2}}^{\text{as1}}\right.\right.\nonumber\\
   &+&\left.\left. 12 f_{\frac{5}{2}}^{\text{as1}}-12 \alpha 
   f_{\frac{3}{2}}^{\ln }+18 \alpha  f_{\frac{5}{2}}^{\ln }+90 f_{\frac{7}{2}}-8
   f_{\frac{3}{2}}^{\ln }+12 f_{\frac{5}{2}}^{\ln }\right)+2 \left(9 \pi ^2
   \left(5 \alpha ^2+8 \alpha +23\right)\right.\right.\nonumber\\
   &+&\left.\left. 8 \left(72 \alpha ^2+365 \alpha
   +442\right)\right) f_{\frac{3}{2}}-9 \left(3 \pi ^2 \left(5 \alpha ^2+8 \alpha
   +23\right)+16 \left(9 \alpha ^2+98 \alpha +160\right)\right) f_{\frac{5}{2}}\right)\nonumber\\
   &-&\frac{2}{3}(\alpha+2)(3\alpha+2)\ln(v \sigma)(2f_{\frac{3}{2}}-3f_{\frac{5}{2}})\big)\Bigg]\,.
\eea

\subsection{The reconstructed field along the source world line: frequency domain computations}

Let us display finally the frequency domain translation of the above result,  i.e., the frequency domain reconstructed field along the source's world line, i.e.,  another original contribution of the present work. Recalling that
\beq
\hat \psi(\omega)=\int dt e^{i \omega t}\psi(t)=\frac{b}{v}\int dT e^{i \hat \omega T}\psi(T)=\frac{b}{v}\hat \psi(\hat\omega)\,,
\eeq
with $\hat \omega=\omega b/v$, we find
\beq
\hat \psi(\hat\omega)=\sum_{n=0}^\infty \hat  \psi^{\epsilon^n}(\hat \omega)\,,
\eeq
where
\bea
\hat  \psi^{\epsilon^0}(\hat \omega)&=&0\,,\nonumber\\
\hat  \psi^{\epsilon^1}(\hat \omega)&=&\frac{q}{b}\Bigg[-2 i (\alpha +1) \hat{\omega } K_0\left(\hat{\omega }\right)\eta^3 v^3+\frac{1}{4} \pi  (\alpha +2) e^{-\hat{\omega }} \left(3 \hat{\omega
   }-1\right)\eta^4v^4\nonumber\\
   &+&\left(\frac{2}{9} i (9 \alpha +14) \hat{\omega }^2 K_1\left(\hat{\omega
   }\right)-\frac{1}{15} i (45 \alpha +83) \hat{\omega }
   K_0\left(\hat{\omega }\right)\right)\eta^5v^5-\frac{1}{8} \pi  (\alpha +2) e^{-\hat{\omega }} \left(5 \hat{\omega
   }^2-10 \hat{\omega }+2\right)\eta^6v^6\Bigg]\,,\nonumber\\
\hat  \psi^{\epsilon^2}(\hat \omega)&=&\frac{q}{b}\Bigg\{-i \pi  (\alpha +1) \hat{\omega }^2 K_0\left(\hat{\omega }\right)\eta^3v^3+\left(\frac{3}{2} (\alpha +2) \hat{\omega }^2 K_0\left(\hat{\omega }\right)-2
   (\alpha +2) \hat{\omega } K_1\left(\hat{\omega }\right)\right.\nonumber\\
   &+&\left.\frac{(\alpha
   +2) e^{-\hat{\omega }} \left(-\pi  \left(2 m_1+3 m_2\right) e^{2
   \hat{\omega }}+2 m_4-3 m_5+2 \pi ^{5/2} \hat{\omega } \left(3
   \hat{\omega }-1\right)\right)}{16 \sqrt{\pi }}\right)\eta^4v^4\nonumber\\
   &+&\left(-\frac{1}{18} i \pi  \left(27 \alpha ^2+99 \alpha +160\right)
   e^{-\hat{\omega }} \hat{\omega }+\frac{1}{30} i \pi  \left(30 \alpha
   ^2+75 \alpha +7\right) \hat{\omega }^2 K_0\left(\hat{\omega
   }\right)+\frac{1}{9} i \pi  (9 \alpha +14) \hat{\omega }^3
   K_1\left(\hat{\omega }\right)\right)\eta^5v^5\nonumber\\
   &+&\Big[-\frac{1}{12} \hat{\omega } \left(8 \alpha +15 \hat{\omega
   }^2+28\right) K_1\left(\hat{\omega }\right)+\frac{1}{64\sqrt{\pi}}\left(\pi  e^{\hat{\omega }} \left(8 (\alpha +2) m_1+4 (3 \alpha +4) m_2-5
   m_3\right)\right.\nonumber\\
   &+&\left.e^{-\hat{\omega }} \left(-8 \pi ^{5/2} (3 \alpha +4) \hat{\omega }^2+8 \pi
   ^{5/2} (\alpha +2) \hat{\omega }-8 (\alpha +2) m_4+4 (3 \alpha +4) m_5+5
   m_6-20 \pi ^{5/2} \hat{\omega }^3\right)\right)\nonumber\\
   &+&K_0\left(\hat{\omega }\right)\left(\frac{2}{3} (3 \alpha +2) \hat{\omega }^2 \ln \left(2 \hat{\omega }\right)+\frac{\hat{\omega}^2}{144}\left(192(2+3\alpha)\ln(v\sigma)+\frac{9\pi^2(23+\alpha(8+5\alpha))}{2+\alpha}\right.\right.\nonumber\\
   &-&\left.\left.96 i \pi  (3 \alpha +2)+96 \gamma  (3 \alpha +2)+8 (3 \alpha +191)\right)\right)\Big](\alpha +2)\eta^6v^6\Bigg\}\,.
\eea
\end{widetext}

Here
\bea
m_1&=&G_{2,3}^{3,0}\left(2 \hat{\omega }|
\begin{array}{c}
 \frac{3}{2},2 \\
 1,1,1 \\
\end{array}
\right)\,,\nonumber\\
m_2&=&G_{2,3}^{3,0}\left(2 \hat{\omega }|
\begin{array}{c}
 \frac{5}{2},3 \\
 2,2,2 \\
\end{array}
\right)\,,\nonumber\\
m_3&=&G_{2,3}^{3,0}\left(2 \hat{\omega }|
\begin{array}{c}
 \frac{7}{2},4 \\
 3,3,3 \\
\end{array}
\right)\,,\nonumber\\
m_4&=&G_{2,3}^{3,1}\left(2 \hat{\omega }|
\begin{array}{c}
 \frac{3}{2},2 \\
 1,1,1 \\
\end{array}
\right)\,,\nonumber\\
m_5&=&G_{2,3}^{3,1}\left(2 \hat{\omega }|
\begin{array}{c}
 \frac{5}{2},3 \\
 2,2,2 \\
\end{array}
\right)\,,\nonumber\\
m_6&=&G_{2,3}^{3,1}\left(2 \hat{\omega }|
\begin{array}{c}
 \frac{7}{2},4 \\
 3,3,3 \\
\end{array}
\right)\,.
\eea
\subsubsection{Soft limit}

In the soft limit $\hat \omega \to 0$ 
\bea
K_0(\hat  \omega)&=& -{\mathcal L} + \frac14 \left(-{\mathcal L}+ 1\right)\hat \omega^2 + O(\hat \omega^4)\,,\nonumber\\
K_1(\hat  \omega)&=& \frac{1}{\hat \omega} + \frac12\left({\mathcal L} - \frac12\right)\hat \omega + O(\hat \omega^3)\,,
\eea
where
\be
\mathcal{L}=\ln\left(\frac{\hat{\omega}e^{\gamma}}{2}\right)\,,
\ee
where, here and below, $\gamma$ denotes the Euler-Mascheroni constant  and 
\begin{widetext}
\bea
\hat \psi^{\epsilon^1}(\hat \omega)&=&\frac{q}{b}\left[2 i (\alpha +1) \hat{\omega } \mathcal{L}\eta^3v^3+\frac{1}{4} \pi  (\alpha +2) \left(4 \hat{\omega }-1\right)\eta^4 v^4+\left(\frac{1}{45} i \hat{\omega } (90 \alpha +3 (45 \alpha +83)
   \mathcal{L}+140)\right.\right.\nonumber\\
   &-&\left.\left.\frac{2}{15} i (45 \alpha +83) \hat{\omega } \ln (2)\right)\eta^5 v^5+\frac{1}{4} \pi  (\alpha +2) \left(6 \hat{\omega }-1\right)\eta^6v^6
\right]+O(\hat \omega^2)\,,\nonumber\\
\hat \psi^{\epsilon^2}(\hat  \omega)&=&\frac{q}{b}\left[ -2 (\alpha +2)\eta^4v^4 -\frac{1}{18} i \pi  (9 \alpha  (3 \alpha +11)+160) \hat{\omega }\eta^5 v^5 -\frac{1}{3} (\alpha +2) (2 \alpha +7)\eta^6 v^6
\right]+O(\hat \omega^2)\,,
\eea
\end{widetext}
where  odd powers of $v$ correspond to purely imaginary contributions.
Not unexpectedly, in the formal limit $\alpha=-2$, i.e. $M_{\rm TS}=0$ (which is physically unacceptable, in fact $r_b=\alpha r_s$ imposes $\alpha>1$ for TS), all the coefficients of $K_0$, $K_1$, $e^{-\hat \omega}$, ${\rm MeijerG}$ denoted as ${\mathcal P}_{K_0}$,
${\mathcal P}_{K_1}$,
${\mathcal P}_{\rm exp}$,
${\mathcal P}_{\rm MeijerG}$  for simplicity (and hence also $\hat \psi^{\epsilon^2,\eta^4}(\hat \omega)$) vanish. This may be taken as a `sanity check' of our formulae.
Note also that
\bea
\hat \psi^{\epsilon^1}(0)&=&\lim_{\hat \omega\to 0}\hat \psi^{\epsilon^1}(\hat \omega)=-\frac{\pi  (\alpha +2) \eta ^4
   q v^4}{4 b}\nonumber\\
   &-&\frac{\pi  (\alpha +2) \eta ^6 q v^6}{4 b}+O(\eta^8)\,,\nonumber\\
\hat \psi^{\epsilon^2}(0)&=&\lim_{\hat \omega\to 0}\hat \psi^{\epsilon^2}(\hat \omega)=-\frac{2 (\alpha +2)
   \eta ^4 q v^4}{b}\nonumber\\
   &-&\frac{(\alpha +2) (2 \alpha +7) \eta ^6 q v^6}{3 b}+O(\eta^8)\,,\qquad
\eea
so that 
\beq
\frac{\hat \psi^{\epsilon^2}(0)}{\hat \psi^{\epsilon^1}(0)}=\frac{8}{\pi}\left(1+\frac{\eta^2 v^2(1+2\alpha)}{6}\right) +O(\eta^4)\,.
\eeq

The $\alpha\to 0$ limit of the above expressions reads
\bea
\hat \psi^{\epsilon^1}(\hat \omega)&=&\frac{q}{b}\left[2 i \eta ^3 v^3 \hat{\omega } \mathcal{L}+\frac{1}{2} \pi  \eta ^4 v^4 \left(4 \hat{\omega }-1\right)\right.\nonumber\\
&+&\left.\eta ^5 \left(\frac{1}{45} i v^5 \hat{\omega } (249
   \mathcal{L}+140)-\frac{166}{15} i v^5 \hat{\omega } \ln (2)\right)\right.\nonumber\\
   &+&\left.\frac{1}{2} \pi  \eta ^6 v^6 \left(6 \hat{\omega }-1\right)
\right]+O(\hat \omega^2)\,,\nonumber\\
\hat \psi^{\epsilon^2}(\hat  \omega)&=&\frac{q}{b}\left[ -4 \eta ^4 v^4 -\frac{80}{9} i \pi  \eta ^5 v^5 \hat{\omega }\right.\nonumber\\
&+&\left.-\frac{14}{3} \eta ^6 v^6
\right]+O(\hat \omega^2)\,,
\eea
and their ratio reduces to
\beq
\lim_{\alpha\to 0}\frac{\hat \psi^{\epsilon^2}(0)}{\hat \psi^{\epsilon^1}(0)}=\frac{8}{\pi}\left(1+\frac{\eta^2 v^2}{6}\right) +O(\eta^4)\,.
\eeq

\begin{figure}
\[
\begin{array}{c}
\includegraphics[scale=0.7]{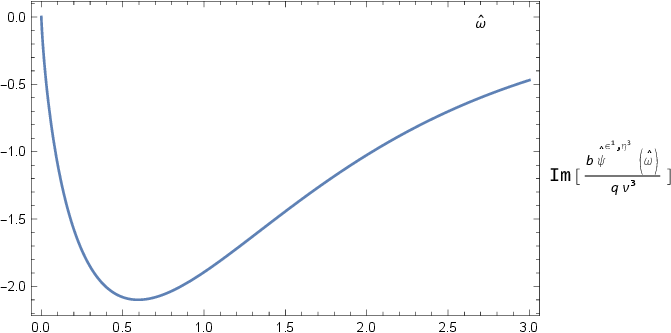} \cr
(a) \cr
 \includegraphics[scale=0.7]{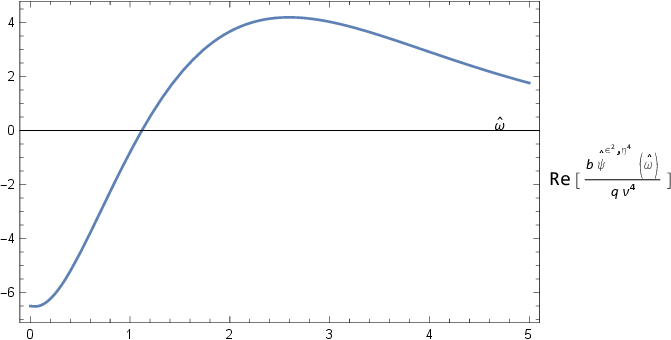} \cr
 (b)\cr
\end{array}
\]
\caption{\label{psi_plots} Examples of plot of the various $\epsilon-\eta$ expanded components of the reconstructed field along the particle's world line for $\alpha=5/4$: (panel a) ${\rm Im}[\frac{b}{qv^3}\hat \psi^{\epsilon^1,\eta^3}(\hat \omega)]$; (panel b) ${\rm Re}[\frac{b}{qv^4}\hat \psi^{\epsilon^2,\eta^4}(\hat \omega)]$.}
\end{figure}

Even if the expression above may appear simple or straightforward, the Fourier transform of the needed integrals is nontrivial and requires a special treatment
for which we follow (and extend) the discussion in Ref. \cite{Bini:2024ijq}, as summarized in the next subsection.
We relegate to an ancillary file all supplemental material containing this kind of (large) expressions, often only in the time domain because of the many  integrals involved and traceable as \lq\lq iterated Bessel functions", see, e.g., Ref. \cite{Bini:2024ijq}.

\subsubsection{Evaluation of the needed integrals}

Most of the integrals appearing in the Fourier transform shown above have been discussed in Ref. \cite{Bini:2024ijq}. Let us derive general expressions for the following classes  
\bea \label{Gint}
Q^{\rm as1}_{n/2,{\tau}}(\hat \omega)&=&\int_{-\infty}^\infty dT e^{i \hat \omega T}\frac{{\rm arcsinh}(\frac{T}{{\tau}})}{({\tau}^2+T^2)^{n/2}}\,,\\
\label{Sint}
Q^{\rm as2}_{n/2,{\tau}}(\hat \omega)&=&\int_{-\infty}^\infty dT e^{i \hat \omega T}\frac{{\rm arcsinh}^2(\frac{T}{{\tau}})}{({\tau}^2+T^2)^{n/2}}\,,\\
\label{S2int}
Q^{\rm as1}_{n,{\tau}}(\hat \omega)&=&\int_{-\infty}^\infty dT e^{i \hat \omega T}\frac{{\rm arcsinh}(\frac{T}{{\tau}})}{({\tau}^2+T^2)^{n}}\,,\\
\label{497}
Q^{\rm as2}_{n,{\tau}}(\hat \omega)&=&\int_{-\infty}^\infty dT e^{i \hat \omega T}\frac{{\rm arcsinh}^2(\frac{T}{{\tau}})}{({\tau}^2+T^2)^{n}}\,.
\eea
where $n$ in the first case is a positive odd integer, while for the last two cases $n$ can be both odd and even. Following Ref. \cite{Bini:2024ijq} let us start with \eqref{Gint}. From the integral identity
\bea
&&\cos(\frac{\nu \pi}{2})K_\nu(x)=\int_0^\infty \cos(x\sinh t)\cosh(\nu t)dt=\nonumber\\
&& =\frac{1}{2}\int_{-\infty}^\infty e^{i x \sinh(t)}\cosh(\nu t)dt \,,
\eea
we can differentiate both sides w.r.t. $\nu$ and introduce the new variable $T=\sinh(t)$
\bea
\label{rel}
&&-\frac{\pi}{2}\sin(\frac{\nu \pi}{2})K_\nu(x)+\cos(\frac{\nu \pi}{2})\frac{d K_\nu(x)}{d\nu}=\nonumber\\
&&\frac{1}{2}\int_{-\infty}^\infty dT e^{i x T}\frac{{\rm arcsinh}(T) \sinh(\nu {\rm arcsinh}(T))}{\sqrt{1+T^2}}\,.
\eea
For $\nu=1$
\be
\int_{-\infty}^\infty e^{i x T}\frac{T {\rm arcsinh}(T)}{\sqrt{1+T^2}}dT=-\pi K_1(x)\,.
\ee
Therefore, integrating over $x$ (with a proper choice of the integration constants) we obtain
\be\label{intGa}
\int_{-\infty}^\infty e^{i x T}\frac{{\rm arcsinh}(T)}{\sqrt{1+T^2}}dT=i \pi K_0(x)\,,
\ee
and from Eq. \eqref{Gint}
\beq
\label{GG1}
Q^{\rm as1}_{1/2,{\tau}}(\hat \omega)=i\pi K_0({\tau} \hat{\omega})\,,
\eeq
we recover the result \eqref{intGa} by setting $\tau=1$.
By taking derivatives w.r.t. ${\tau}$, we can find the following recursive formula
\be\label{recG}
Q^{\rm as1}_{\frac{n+2}{2},{\tau}}(\hat \omega)=-\frac{1}{\tau n} \partial_\tau Q^{\rm as1}_{\frac{n}{2},{\tau}}(\hat \omega)-\frac{i 2^{1-\frac{n}{2}}\sqrt{\pi}\hat{\omega}^{\frac{n}{2}}K_{1-\frac{n}{2}}({\tau} \hat{\omega})}{{\tau}^{1+\frac{n}{2}}n \Gamma(\frac{n+1}{2})}\,,
\ee
and then, starting from \eqref{GG1} and using \eqref{recG}, we can determine recursively all the other integrals.
Let us now compute the integral shown in Eq. \eqref{Sint}. From \cite{Bini:2024ijq,bry2016} we can write:
\be
Q^{\rm as2}_{1/2,{\tau}}(\hat \omega){=}{-}\frac{\pi^2}{2}K_0(\tau \hat{\omega}){+}2\frac{d^2K_\nu(\tau \hat{\omega})}{d\nu^2}|_{\nu=0}\,,
\ee
where
\begin{widetext}
  \bea
\frac{d^2K_\nu(\tau \hat{\omega})}{d\nu^2}|_{\nu=0}&=&A_{K_0}K_0(\tau\hat{\omega})+B_{I_0}I_0(\tau\hat{\omega})\,,\nonumber\\
A_{K_0}&=&\frac{\pi ^2}{2}{-}\frac{1}{4} i \pi  \tau^2\hat{\omega}^2 \, _3F_4\left(1,1,\frac{3}{2};2,2,2,2;\tau^2\hat{\omega}^2\right){-}\frac{1}{2} \sqrt{\pi } G_{3,5}^{3,1}\left(-\tau^2\hat{\omega}^2|
\begin{array}{c}
 \frac{1}{2},-\frac{1}{2},1 \\
 0,0,0,-\frac{1}{2},0 \\
\end{array}
\right){-}i \pi  \ln \left(\frac{\tau\hat{\omega}}{2}e^{\gamma} \right)\,,\nonumber\\
B_{I_0}&=&-\frac{1}{2} \pi ^{3/2} G_{3,5}^{4,0}\left(-\tau^2\hat{\omega}^2|
\begin{array}{c}
 \frac{1}{2},\frac{1}{2},1 \\
 0,0,0,0,\frac{1}{2} \\
\end{array}
\right)+\frac{1}{2} i \pi ^{3/2} G_{3,5}^{3,1}\left(-\tau^2\hat{\omega}^2|
\begin{array}{c}
 \frac{1}{2},-\frac{1}{2},1 \\
 0,0,0,-\frac{1}{2},0 \\
\end{array}
\right)\,,
\eea  
\end{widetext}
where $_3F_4$ is a generalized hypergeometric function and $G_{p,q}^{k,r}$ are Meijer $G$ functions.
 
All the integrals of this family can be computed recursively by using the recurrence relation
\beq
Q^{as2}_{\frac{n}{2}+1,\tau}=\frac{2i}{\tau^2 n}\partial_{\hat{\omega}}Q^{as1}_{\frac{n+1}{2},\tau}-\frac{1}{n \tau}\partial_\tau Q^{as2}_{\frac{n}{2},\tau}\,.
\eeq
Some of values of these integrals are computed in table \ref{tabas2semint} and compared with numerical integration.
\begin{table}[]
\begin{tabular}{|c|c|c|}
\hline
n       & Recursion                  & Numerical                  \\ \hline
1       & $-1.4620502381291095405$ & $-1.4620502381(4)$  \\ \hline
3       & $-0.030925267261887709292$ & $-0.030925267261887709292$ \\ \hline
5       & $0.14893783065003824060$ & $0.148937830650038240(3)$ \\ \hline
7       & $0.126031900636258113378$ & $0.126031900636(3)$ \\ \hline
9       & $0.094266170107469157055$ & $0.09426617010746915705(6)$ \\ \hline
11       & $0.071617515853924140279$ & $0.071617515853924140279$ \\ \hline
\end{tabular}
\caption{\label{tabas2semint} Comparison of the values obtained with the recursion formula for $Q^{\rm as2}_{n/2,\tau}(\hat \omega)$ and the numerical integration of \eqref{Sint} for $\tau=\hat{\omega}=1$.}
\end{table}
Concerning Eq. \eqref{S2int}, the only integral that we need to compute is
\bea
&&Q^{\rm as1}_{1,{\tau}}(\hat \omega)=\int_{-\infty}^\infty e^{i \hat \omega T}\frac{{\rm arcsinh}(\frac{T}{{\tau}})}{{\tau}^2+T^2}=\nonumber\\
&& \frac{i}{{\tau}}\Bigg[\frac{\pi^2}{2}e^{-{\tau} \hat \omega}+e^{{\tau} \hat \omega}\sqrt{\pi}G_{2,3}^{3,0}\left(2\tau \hat \omega \bigg|\begin{array}{c} \frac{1}{2},1\\ 0,0,0 \end{array} \right)-\nonumber\\
&&\frac{e^{-{\tau} \hat \omega}}{\sqrt{\pi}}G_{2,0}^{3,1}\left(2{\tau} \hat \omega \bigg|\begin{array}{c} \frac{1}{2},1\\ 0,0,0 \end{array} \right)\Bigg]\,.
\eea
Starting from the previous result, all the other integrals of the family follow by using the recurrence relation
\bea
Q^{\rm as1}_{n+1,\tau}(\hat \omega) &=& -\frac{1}{2\tau n} \partial_\tau Q^{\rm as1}_{n,\tau}(\hat \omega) \nonumber\\
&-& \frac{i 2^{-n} \sqrt{\pi} \hat \omega^n \tau^{-1-n} K_{n-1}(\tau \hat \omega)}{n \Gamma\left(n + \frac{1}{2}\right)}\,.\qquad
\eea
The analytical results are always checked against numerical evaluations, as shown in Tables \ref{tab1} and \ref{tab2} below.

\begin{table}[]
\begin{tabular}{|c|c|c|}
\hline
n       & Recursion                  & Numerical                  \\ \hline
1 & $1.3226872821587789124 \, i$  & $1.32268728215(0)\, i$  \\ \hline
3       & $0.73521998273929097756 \, i$ & $0.73521998273(7) \, i$ \\ \hline
5       & $0.35317874098365943022 \, i$ & $0.353178740983659(6)\, i$ \\ \hline
7       & $0.19672280082727274226 \, i$ & $0.19672280082727(3)  \, i$ \\ \hline
\end{tabular}
\caption{\label{tab1} Comparison of the values obtained with the recursion formula for $Q^{\rm as1}_{n/2,\tau}(\hat \omega)$ and the numerical integration of \eqref{Gint} for $\tau=\hat{\omega}=1$.}
\end{table}

\begin{table}[]
\begin{tabular}{|c|c|c|}
\hline
n       & Recursion                  & Numerical                  \\ \hline
1       & $1.0581762416658229435 \, i$  & $1.0581762416(3)  \, i$  \\ \hline
2       & $0.50256820218299371586 \, i$ & $0.5025682021829(7) \, i$ \\ \hline
3       & $0.25840350259662541536 \, i$ & $0.258403502596625(6) \, i$ \\ \hline
\end{tabular}
\caption{\label{tab2} Comparison of the values obtained with the recursion formula for $Q^{\rm as1}_{n,\tau}(\hat \omega)$ and the numerical integration of \eqref{S2int} for $\tau=\hat{\omega}=1$.}
\end{table}

The next integral needed is 
\beq
\label{Q_as2_1}
Q^{\rm as2}_1(\hat \omega) =\int dT e^{i\hat \omega T}\frac{{\rm arcsinh}^2(T)}{1+T^2}\,,
\eeq
which is such that
\bea
\label{gen_eq_Q_as2}
Q^{\rm as2}_1(\hat \omega)''-Q^{\rm as2}_1(\hat \omega) &=&  -\int dT e^{i\hat \omega T} {\rm arcsinh}^2(T) \nonumber\\
&=& \frac{2\pi}{\hat \omega} K_0(\hat \omega)\,,
\eea
and hence  it is included in the class of iterated Bessel functions.
From Eq. \eqref{Q_as2_1} we find
\beq
\label{Q_as2_1_0}
Q^{\rm as2}_1(0) =\int dT \frac{{\rm arcsinh}^2(T)}{T^2+1}=\frac{\pi^3}{4}\,.
\eeq 

The general solution of Eq. \eqref{gen_eq_Q_as2}, 
\bea
Q^{\rm as2}_1(\hat \omega)&=& -\sqrt{\pi } e^{-\hat \omega} \left(\pi  e^{2 \hat \omega} G_{2,3}^{3,0}\left(2 \hat \omega\left|
\begin{array}{c}
 \frac{1}{2},1 \\
 0,0,0 \\
\end{array}
\right.\right)\right.\nonumber\\
&-&\left. G_{2,3}^{3,1}\left(2 \hat \omega\left|
\begin{array}{c}
 \frac{1}{2},1 \\
 0,0,0 \\
\end{array}
\right.\right)\right)\nonumber\\
&+& c_1 e^{\hat \omega}+c_2 e^{-\hat \omega}\,,
\eea
with $c_1$ and $c_2$ determined as follows.
Evaluating this expression in $\hat \omega=0$ gives
\beq
\frac{\pi^3}{2} + c_1+c_2=\frac{\pi^3}{4}\qquad \to\qquad c_1+c_2=-\frac{\pi^3}{4}\,.
\eeq
Inspecting the behavior of the (real part of the) integral \eqref{Q_as2_1} for large (and positive) $\hat \omega$ we see that it vanishes. Therefore we can choose
\beq
c_1=0\,.
\eeq
Finally,
\bea
Q^{\rm as2}_1(\hat \omega)&=& -\sqrt{\pi } e^{-\hat \omega} \left(\pi  e^{2 \hat \omega} G_{2,3}^{3,0}\left(2 \hat \omega\left|
\begin{array}{c}
 \frac{1}{2},1 \\
 0,0,0 \\
\end{array}
\right.\right)\right.\nonumber\\
&-&\left. G_{2,3}^{3,1}\left(2 \hat \omega\left|
\begin{array}{c}
 \frac{1}{2},1 \\
 0,0,0 \\
\end{array}
\right.\right)\right)\nonumber\\
&-&\frac{\pi^3}{4} e^{-\hat \omega}\,,
\eea
which agrees with values obtained  by integrating numerically the defining integral.
Note that this result uses the general methodology outlined in Ref. \cite{Bini:2024ijq}  but was not derived previously.

In order to obtain the other integrals of the family, let us define 
\bea
Q^{\rm as2}_{n,\tau}(\hat \omega)&=&\int_{-\infty}^\infty dT \frac{e^{i \hat{\omega} T}}{(\tau^2+T^2)^n}{\rm arcsinh}^2\left(\frac{T}{\tau}\right)\nonumber\\
&=&\tau^{1-2n}Q^{\rm as2}_{n}(\tau \hat \omega)\,.
\eea
Taking a derivative w.r.t. $\tau$ yields to
\beq
Q^{\rm as2}_{n+1,\tau}(\hat \omega)=\frac{i}{n \tau^2}\partial_{\hat \omega} Q^{\rm as1}_{n+1/2,\tau}(\hat \omega)-\frac{1}{2 n \tau}\partial_\tau Q^{\rm as2}_{n,\tau}(\hat \omega)\,.
\eeq
In table \ref{tabas2} we check numerically the results of the recursion relations.
\begin{table}[]
\begin{tabular}{|c|c|c|}
\hline
n       & Recursion                  & Numerical                  \\ \hline
1       & $-0.47271577253165162866$  & $-0.4727157725316(7)$  \\ \hline
2       & $0.11705101279211073944$ & $0.117051012792110739(3)$ \\ \hline
3       & $0.14241922253717755533$ & $0.14241922253717755(7)$ \\ \hline
4       & $0.10911919339039772457$ & $0.10911919339039772(8)$ \\ \hline
\end{tabular}
\caption{\label{tabas2} Comparison of the values obtained with the recursion formula for $Q^{\rm as2}_{n,\tau}(\hat \omega)$ and the numerical integration of \eqref{497} for $\tau=\hat{\omega}=1$.}
\end{table}

The next integral to consider is
\beq
Q^{\rm as3}_{1/2}(x)=\int dT e^{ixT}\frac{{\rm arcsinh}^3(T)}{\sqrt{1+T^2}}
\eeq
and following the same procedure outlined above one can show that
\beq
Q^{\rm as3}_{1/2}(x)=\frac{\pi^3}{4}K_1(x)-3\pi \partial_{\nu}^2K_\nu(x)\bigg|_{\nu=1}\,,
\eeq
where $\partial_\nu^2 K_\nu(x) \bigg|_{\nu=1}$ can be found, for example, in Eqs. B2-B5 of Ref.  \cite{Bini:2024ijq}.
Generalizing as before,
\beq
Q^{\rm as3}_{n,{\tau}}(x)=\int dT e^{ixT}\frac{{\rm arcsinh}^3(\frac{T}{{\tau}})}{({\tau}^2+T^2)^n}\,,
\eeq
so that
\beq
Q^{\rm as3}_{n+1,{\tau}}(x)=-\frac{1}{2n {\tau} }\partial_\tau Q^{\rm as3}_{n,{\tau}}(x) +\frac{3i}{2n{\tau}^2} \partial_x Q^{\rm as2}_{n+\frac12,{\tau}}(x)\,,
\eeq
which can be used for the recursion.

Finally, including higher powers of arcsinh
\beq
Q^{\rm as4}_{0}(x)=\int dT e^{ixT} {\rm arcsinh}^4(T)\,,
\eeq
one has
\bea
Q^{\rm as4}_{0}(x)&=& \pi^3\partial_\nu K_\nu(x)\bigg|_{\nu=0}-4\pi \partial^3_{\nu} K_\nu(x)\bigg|_{\nu=1}\nonumber\\
&=& -4\pi \partial^3_{\nu}K_\nu(x)\bigg|_{\nu=1}\,,
\eea 
where $\partial^3_{\nu}K_\nu(x)\bigg|_{\nu=1}$ is given in Ref. \cite{bry2016} (and it is the last explicitly known expression of derivatives of the Bessel K functions with respect to the order).

\section{Scalar self force}
In $D=4$ the definition of scalar self force along the source's world line reads
\beq
\label{SF_def}
F_\alpha(\tau) =q P(u)_\alpha{}^\beta \partial_\beta \psi \bigg|_{x^\alpha=x^\alpha(\tau)}\,,
\eeq 
where $P(u)_\alpha{}^\beta=\delta_\alpha^\beta +u_\alpha u^\beta$ 
projects orthogonally to $u^\alpha$, the four velocity of the source.
In general, we will distinguish left/right force components, $F_\alpha^\pm(\tau)$, because the field is continuous across the source world line but not its derivatives.

The field at a generic spacetime point has been given in  Eq. \eqref{psi_val_TS_1}. Thanks to spherical symmetry, both geodetic and non-geodetic motion can be taken to be on the equatorial plane $\theta=\frac{\pi}{2}$ so that $F_\theta=0$. Moreover, due to the orthogonality condition $F_\alpha u^\alpha=0$, only $F_t$ and $F_\phi$ should be computed 
since
\bea
F_r&=& -\frac{u^t}{u^r}\left(F_t  +\frac{u^\phi}{u^t}F_\phi\right)\nonumber\\
&=&  -\left(\frac{dr_p}{dt}\right)^{-1}\left(F_t  +\frac{d\phi_p}{dt}F_\phi\right)\,,
\eea
with
\bea
F_t &=& q (\partial_t \psi +u_t u^\beta \partial_\beta \psi)\,,\nonumber\\
F_\phi &=& q (\partial_\phi \psi +u_\phi u^\beta \partial_\beta \psi)\,,
\eea
where now we have to consider left and right parts of the field separately (since derivatives may be discontinuous along the world line)
\beq
\psi^\pm(t,r,\theta,\phi)=\sum_{lm}Y_{lm}(\theta,\phi)\int \frac{d\omega}{2\pi}e^{-i\omega t}   R_{l{}\omega}^\pm(r)\,,
\eeq 
with
\bea
\partial_t \psi^\pm(t,r,\theta,\phi)&=&\sum_{lm}Y_{lm}(\theta,\phi)\int \frac{d\omega}{2\pi}e^{-i\omega t}  (-i\omega) R_{l{}\omega}^\pm(r)\,,\nonumber\\
\partial_\phi \psi^\pm(t,r,\theta,\phi)&=&\sum_{lm}(im) Y_{lm}(\theta,\phi)\int \frac{d\omega}{2\pi}e^{-i\omega t}  R_{l{}\omega}^\pm(r)\,,\nonumber\\
\partial_r \psi^\pm(t,r,\theta,\phi)&=&\sum_{lm}Y_{lm}(\theta,\phi)\int \frac{d\omega}{2\pi}e^{-i\omega t}\partial_r R_{l{}\omega}^\pm(r)\,,
\eea
and then define a corresponding left and right force, $F_\alpha^\pm$, evaluated along the orbit, using
\bea
&&\partial_t \psi^\pm(t,r,\theta,\phi)\big|_{\rm orb}\,,\quad 
\partial_\phi \psi^\pm(t,r,\theta,\phi)\big|_{\rm orb}\,,\nonumber\\
&& \partial_r \psi^\pm(t,r,\theta,\phi)\big|_{\rm orb}\,.
\eea
Let us recall here Eq. \eqref{R_pm}, that is
\bea
\label{R_pm_bis}
R_{lm\omega}^\pm (r)&=&-4\pi q Y^*_{lm}(\frac{\pi}{2},0) \int dt' G_{l{}\omega}^\pm (r,r_p(t'))\times \nonumber\\
&& \frac{\sqrt{f_b(r_p(t'))}e^{i(\omega t'-m\phi_p(t'))}  }{u^t(t')}\,,
\eea 
with
\bea
G_{l{}\omega}^+ (r,r')&=& \frac{R_{\rm in}^{l{}\omega}(r')R_{\rm up}^{l{}\omega}(r)}{W_{l{}\omega}}\,,\nonumber\\
G_{l{}\omega}^- (r,r')&=& \frac{R_{\rm in}^{l{}\omega}(r)R_{\rm up}^{l{}\omega}(r')}{W_{l{}\omega}} \,.
\eea
We find it convenient to integrate over $\omega$ first and then over $t'$. Therefore, for example
\begin{widetext}
\bea
\partial_t \psi^\pm(t,r,\theta,\phi)\bigg|_{\rm orb}&=&-4\pi q  \sum_{lm}|Y_{lm}(\frac{\pi}{2},0)|^2\int dt' \int \frac{d\omega}{2\pi}e^{-i\omega (t-t')}  (-i\omega) G_{l{}\omega}^\pm (r_p(t),r_p(t'))  \frac{\sqrt{f_b(r_p(t'))}e^{i m(\phi_p(t)-\phi_p(t'))}  }{u^t(t')}\nonumber\\
&=&-4\pi q  \sum_{lm}|Y_{lm}(\frac{\pi}{2},0)|^2\int dt' \, \frac{\sqrt{f_b(r_p(t'))}e^{i m(\phi_p(t)-\phi_p(t'))}  }{u^t(t')}\, {\rm FT}[(-i\omega) G_{l{}\omega}^\pm (r_p(t),r_p(t')) ]\,,
\eea
where
\beq
{\rm FT}[f(\omega, t,t')]=\int \frac{d\omega}{2\pi}e^{-i\omega (t-t')}f(\omega, t,t')\,,
\eeq
and
\bea
\partial_\phi \psi^\pm(t,r,\theta,\phi)\bigg|_{\rm orb}
&=&-4\pi q  \sum_{lm}|Y_{lm}(\frac{\pi}{2},0)|^2(im)\int dt' \, \frac{\sqrt{f_b(r_p(t'))}e^{i m(\phi_p(t)-\phi_p(t'))}  }{u^t(t')}\, {\rm FT}[ G_{l{}\omega}^\pm (r_p(t),r_p(t')) ]\,,\nonumber\\
\partial_r \psi^\pm(t,r,\theta,\phi)\bigg|_{\rm orb}
&=&-4\pi q  \sum_{lm}|Y_{lm}(\frac{\pi}{2},0)|^2 \int dt' \, \frac{\sqrt{f_b(r_p(t'))}e^{i m(\phi_p(t)-\phi_p(t'))}  }{u^t(t')}\, {\rm FT}[ \partial_r G_{l{}\omega}^\pm (r,r_p(t'))|_{r=r_p(t)} ]\,.\nonumber\\
\eea

\end{widetext}

Once the derivatives of $\psi$ have been taken and  restricted to the orbit one can define the left and right  components of the force (along the particle's world line),
\bea
F_t^\pm&=& q[(1+u_tu^t)\partial_t \psi^\pm+u_tu^r\partial_r \psi^\pm+u_tu^\phi \partial_\phi \psi^\pm]\,,\nonumber\\
F_r^\pm&=& q[ u_ru^t \partial_t \psi^\pm+(1+u_ru^r)\partial_r \psi^\pm+u_r u^\phi \partial_\phi \psi^\pm]\,,\nonumber\\
F_\phi^\pm&=& q[ u_\phi u^t \partial_t \psi^\pm+u_\phi u^r\partial_r \psi^\pm+(1+u_\phi u^\phi) \partial_\phi \psi^\pm]\,.\nonumber\\
\eea
After manipulations, for example
\bea
F_t^\pm&=& q\left((1-\kappa)\partial_t \psi^\pm-\kappa  \frac{dr_p}{dt}\partial_r \psi^\pm-\kappa \frac{d\phi_p}{dt}\partial_\phi \psi^\pm\right)\,,\qquad
\eea
where
\bea
\kappa&=&f_s(r_p(t)) (u^t)^2\nonumber\\
&=& \kappa^{\epsilon^0}+\epsilon \kappa^{\epsilon^1}+\epsilon^2\kappa^{\epsilon^2}+O(\epsilon^3)\,,
\eea
with
\bea
\kappa^{\epsilon^0}&=&\gamma^2=1+v^2\eta^2+ v^4\eta^4+v^6\eta^6+O(\eta^8) \nonumber\\ 
\kappa^{\epsilon^1}&=&\frac{2\left(v^2\eta^2+v^4\eta^4+v^6\eta^6\right)}{\sqrt{1+T^2}}+O(\eta^8) \nonumber\\ 
\kappa^{\epsilon^2}&=& \frac{2 \eta ^2 v^2 \left(\sqrt{T^2+1}{-}T {\rm arcsinh}(T)\right)}{\left(T^2{+}1\right)^{3/2}}\nonumber\\
   &{+}&\frac{2 \eta ^4 v^4 \left(3 \sqrt{T^2{+}1}{+}(\alpha {+}2)
   T {\rm arcsinh}(T)\right)}{\left(T^2{+}1\right)^{3/2}}\nonumber\\
   &{+}&\frac{2 \eta ^6 v^6 \left(3 \sqrt{T^2{+}1}{+}(\alpha {+}2)
   T {\rm arcsinh}(T)\right)}{\left(T^2{+}1\right)^{3/2}}\,.
\eea
One then regularizes the results removing all jumps across the source's world line
\beq
F_\alpha^{\rm reg}\bigg|_{\rm orb}=\frac12 (F_\alpha^++F_\alpha^-)\bigg|_{\rm orb}\,.
\eeq
This process leads to an $F_\alpha^{\rm reg}(T)$. In addition, the regularized force components split in a conservative and a dissipative parts
\bea
F_\alpha^{\rm reg, cons}(T)&=& \frac12 \left(F_\alpha^{\rm reg}(T)-F_\alpha^{\rm reg}(-T) \right)\,,\nonumber\\
F_\alpha^{\rm reg, diss}(T)&=& \frac12 \left(F_\alpha^{\rm reg}(T)+F_\alpha^{\rm reg}(-T) \right)\,,
\eea
for $\alpha=t,\phi$, while the signs are opposite for the $r$ component. We will compute the components of the (regularized) self force, and distinguish among conservative and dissipative contributions,  
\bea
&& F_t^{\rm reg, cons}(T)\,,\quad F_r^{\rm reg, cons}(T)\,,\quad F_\phi^{\rm reg, cons}(T)\,,\nonumber\\
&& F_t^{\rm reg, diss}(T)\,,\quad F_r^{\rm reg, diss}(T)\,,\quad F_\phi^{\rm reg, diss}(T)\,.
\eea
Our findings for both the derivatives of the field and the components of the self force are summarized in the following tables: \ref{dt_psi}, 
\ref{dt_psi_diss}, 
\ref{F_t}.

\vspace{1cm}

\begin{widetext}

\begin{table}[]
\caption{PM-PN expansion of the conservative part of $\partial_t \psi$, $\partial_r \psi$, $\partial_\phi \psi$.}\label{dt_psi}
\begin{ruledtabular}
\begin{tabular}{|c|l|}
\hline
$\partial_t \psi_{\rm reg, cons}^{\eta^4, \epsilon^1}$                       & $\frac{(\alpha +2) q v^4 \left(2 \tilde{f}_2-5 \tilde{f}_3\right)}{b}$ \\ \hline
$\partial_t \psi_{\rm reg, cons}^{\eta^6, \epsilon^1}$                       & $\frac{(\alpha +2) q v^6 \left(34 \tilde{f}_2-205 \tilde{f}_3+210
   \tilde{f}_4\right)}{6 b}$ \\ \hline
$\partial_t \psi_{\rm reg, cons}^{\eta^4, \epsilon^2}$                       & $-\frac{(\alpha +2) q v^4 \left(-17 \tilde{f}_{\frac{5}{2}}+30
   \tilde{f}_{\frac{7}{2}}+6 f_2^{\text{as1}}-33 f_3^{\text{as1}}+30
   f_4^{\text{as1}}\right)}{b}$\\  \hline
$\partial_t \psi_{\rm reg, cons}^{\eta^6, \epsilon^2}$                       & $\frac{q v^6}{192 b} \Big[3 \alpha ^2 \left(6 (256 \gamma +\pi  (9 \pi -128 i)+256 (\ln
   (2)-3)) \tilde{f}_{\frac{5}{2}}+(9952-3840 \gamma +15 (-9 \pi +128 i) \pi -3840
   \ln (2)) \tilde{f}_{\frac{7}{2}}\right.$ \\
   & $\left.+1152 \left(2 \tilde{f}_{\frac{5}{2}}^{\ln }-5
   \tilde{f}_{\frac{7}{2}}^{\ln }\right)-2112 f_3^{\text{as1}}+1920
   f_4^{\text{as1}}\right)+4 \alpha  \left(2 \left(-2404+1344 \gamma -672 i \pi +27
   \pi ^2+1344 \ln (2)\right) \tilde{f}_{\frac{5}{2}}\right.$\\ 
   & $\left.+(2816-6720 \gamma +15 (-9 \pi
   +224 i) \pi -6720 \ln (2)) \tilde{f}_{\frac{7}{2}}+672 \left(20
   \tilde{f}_{\frac{9}{2}}+6 \tilde{f}_{\frac{5}{2}}^{\ln }-15
   \tilde{f}_{\frac{7}{2}}^{\ln }\right)+1368 f_3^{\text{as1}}-14400
   f_4^{\text{as1}}\right.$ \\
   & $\left.+13440 f_5^{\text{as1}}\right)+2 \left(\left(8416+1536 \gamma
   -768 i \pi +351 \pi ^2\right) \tilde{f}_{\frac{5}{2}}+64 \left(-757
   \tilde{f}_{\frac{7}{2}}+840 \tilde{f}_{\frac{9}{2}}+36
   \tilde{f}_{\frac{5}{2}}^{\ln }-90 \tilde{f}_{\frac{7}{2}}^{\ln }\right)+18144
   f_3^{\text{as1}}\right.$ \\
   & $\left.-69120 f_4^{\text{as1}}+53760 f_5^{\text{as1}}\right)+384 (\alpha
   +2) (3 \alpha +1) \left(2 \tilde{f}_{\frac{5}{2}}-5
   \tilde{f}_{\frac{7}{2}}\right) \left(\ln (v)-\ln \left(\sigma
   _t\right)\right)+3072 \ln (2) \tilde{f}_{\frac{5}{2}}$\\
   & $-15 (512 \gamma +\pi  (117
   \pi -256 i)+512 \ln (2)) \tilde{f}_{\frac{7}{2}}+192 (\alpha +2) (6 \alpha +1)
   f_2^{\text{as1}}\Big]$ \\\hline
\hline
$\partial_r \psi_{\rm reg, cons}^{\eta^6, \epsilon^1}$                       & $-\frac{(\alpha +2) \left(8 f_{\frac{3}{2}}-40 f_{\frac{5}{2}}+35 f_{\frac{7}{2}}\right) q v^6}{3 b^2}$ \\ \hline
$\partial_r \psi_{\rm reg, cons}^{\eta^6, \epsilon^2}$                       & $\frac{q v^6}{192 b^2} \Big[64 (\alpha +2) \left(24 \tilde{f}_{\frac{5}{2}}^{\text{as1}}-200 \tilde{f}_{\frac{7}{2}}^{\text{as1}}+245 \tilde{f}_{\frac{9}{2}}^{\text{as1}}-245 f_4-12
   f_2^{\ln }+18 f_3^{\ln }\right)$ \\
   & $+2 \left(9 \pi ^2 (\alpha  (\alpha +4)+7)-64 (\alpha +2) (3 \alpha +41)\right) f_2+\left(64 (\alpha +2) (9 \alpha +313)-27 \pi ^2 (\alpha 
   (\alpha +4)+7)\right) f_3$\\ 
   & $+384 (\alpha +2) \left(2 f_2-3 f_3\right) \ln \left(v \sigma _r\right)\Big]$ \\\hline
\hline
$\partial_\phi \psi_{\rm reg, cons}^{\eta^4, \epsilon^1}$                       & $-\frac{(\alpha +2) q v^4 \tilde{f}_2}{b}$ \\ \hline
$\partial_\phi \psi_{\rm reg, cons}^{\eta^4, \epsilon^2}$                       & $\frac{(\alpha +2) q v^4 \left(3 f_2^{\text{as1}}-4
   \left(\tilde{f}_{\frac{5}{2}}+f_3^{\text{as1}}\right)\right)}{b}$ \\ \hline
$\partial_\phi \psi_{\rm reg, cons}^{\eta^6, \epsilon^1}$ & $-\frac{5 (\alpha +2) q v^6 \left(5 \tilde{f}_2-8 \tilde{f}_3\right)}{6 b}$ \\\hline
$\partial_\phi \psi_{\rm reg, cons}^{\eta^6, \epsilon^2}$ & $\frac{q v^6}{96 b} \Big[\alpha  \left(-\left(32 (68-9 \alpha )+9 \pi ^2 (\alpha +4)\right)
   \tilde{f}_{\frac{5}{2}}+128 \tilde{f}_{\frac{3}{2}}+384 \left(10
   \tilde{f}_{\frac{7}{2}}+\tilde{f}_{\frac{5}{2}}^{\ln }\right)+3840
   f_4^{\text{as1}}\right)-384 (\alpha +2) \tilde{f}_{\frac{5}{2}} \ln
   (\sigma_\phi v)$\\
   & $+256 \tilde{f}_{\frac{3}{2}}-\left(5504+63 \pi ^2\right)
   \tilde{f}_{\frac{5}{2}}+768 \left(10
   \tilde{f}_{\frac{7}{2}}+\tilde{f}_{\frac{5}{2}}^{\ln }\right)-48 (\alpha +2) (6
   \alpha -7) f_2^{\text{as1}}+192 (\alpha +2) (2 \alpha -19) f_3^{\text{as1}}+7680
   f_4^{\text{as1}}\Big]$ \\\hline
\end{tabular}
\end{ruledtabular}
\end{table}

\begin{table}[]
\caption{PM-PN expansion of the dissipative part of $\partial_t \psi$, $\partial_r \psi$, $\partial_\phi \psi$.}\label{dt_psi_diss}
\begin{ruledtabular}
\begin{tabular}{|c|l|}
\hline
$\partial_t \psi_{\rm reg, diss}^{\eta^3, \epsilon^1}$                       & $\frac{(3 \alpha +2) \left(2 f_{\frac{3}{2}}-3 f_{\frac{5}{2}}\right) q v^3}{3 b}$ \\ \hline
$\partial_t \psi_{\rm reg, diss}^{\eta^3, \epsilon^2}$                       & $\frac{q v^3 \left((33 \alpha +23) f_2-3 (3 \alpha +2) \left(2
   \tilde{f}_{\frac{5}{2}}^{\text{as1}}-5 \tilde{f}_{\frac{7}{2}}^{\text{as1}}+5
   f_3\right)\right)}{3 b}$ \\ \hline
$\partial_t \psi_{\rm reg, diss}^{\eta^5, \epsilon^1}$                       & $\frac{q v^5 \left(2 (85 \alpha +126) f_{\frac{3}{2}}-(615 \alpha +778)
   f_{\frac{5}{2}}+50 (9 \alpha +10) f_{\frac{7}{2}}\right)}{30 b}$ \\ \hline
$\partial_t \psi_{\rm reg, diss}^{\eta^5, \epsilon^2}$                       & $\frac{q v^5}{90 b} \Big[3 \left(15 \alpha  \left(2 (6 \alpha +5)
   \tilde{f}_{\frac{5}{2}}^{\text{as1}}+5 (19-6 \alpha )
   \tilde{f}_{\frac{7}{2}}^{\text{as1}}-210
   \tilde{f}_{\frac{9}{2}}^{\text{as1}}\right)-396
   \tilde{f}_{\frac{5}{2}}^{\text{as1}}+2990
   \tilde{f}_{\frac{7}{2}}^{\text{as1}}-3500 \tilde{f}_{\frac{9}{2}}^{\text{as1}}+5
   (\alpha  (18 \alpha -815)\right.$ \\ 
 & $\left.-1310) f_3+350 (9 \alpha +10) f_4\right)+(15 (221-12
   \alpha ) \alpha +8149) f_2\Big]$\\\hline
$\partial_t \psi_{\rm reg, diss}^{\eta^6, \epsilon^2}$                       & $-\frac{2 (\alpha +2) (3 \alpha +1) q v^6 \left(-6
   \tilde{f}_{\frac{5}{2}}^{\text{as1}}+15 \tilde{f}_{\frac{7}{2}}^{\text{as1}}+11
   f_2-15 f_3\right)}{3 b}$ \\\hline
\hline
$\partial_r \psi_{\rm reg, diss}^{\eta^3, \epsilon^1}$                       & $\frac{2 q v^3 \tilde{f}_2}{3 b^2}$ \\ \hline
$\partial_r \psi_{\rm reg, diss}^{\eta^3, \epsilon^2}$                       & $\frac{2 q v^3 \left(4 \left(\tilde{f}_{\frac{5}{2}}+f_3^{\text{as1}}\right)-3
   f_2^{\text{as1}}\right)}{3 b^2}$ \\ \hline
$\partial_r \psi_{\rm reg, diss}^{\eta^5, \epsilon^1}$                       & $\frac{q v^5 \left(25 (3 \alpha +2) \tilde{f}_3-(40 \alpha +43) \tilde{f}_2\right)}{15
   b^2}$ \\ \hline
$\partial_r \psi_{\rm reg, diss}^{\eta^5, \epsilon^2}$                       & $\frac{q v^5 \left(-990 \alpha  \tilde{f}_{\frac{5}{2}}+450 (3 \alpha +2)
   \tilde{f}_{\frac{7}{2}}-1151 \tilde{f}_{\frac{5}{2}}+9 (50 \alpha +73)
   f_2^{\text{as1}}-3 (575 \alpha +542) f_3^{\text{as1}}+1350 \alpha 
   f_4^{\text{as1}}+900 f_4^{\text{as1}}\right)}{45 b^2}$ \\  \hline
$\partial_r \psi_{\rm reg, diss}^{\eta^6, \epsilon^2}$ & $-\frac{2 (\alpha +2) q v^6 \left(2 f_2^{\text{as1}}-3
   \left(\tilde{f}_{\frac{5}{2}}+f_3^{\text{as1}}\right)\right)}{b^2}$ \\\hline
\hline
$\partial_\phi \psi_{\rm reg, diss}^{\eta^3, \epsilon^1}$                       & $-\frac{f_{\frac{3}{2}} q v^3}{3 b}$ \\ \hline
$\partial_\phi \psi_{\rm reg, diss}^{\eta^3, \epsilon^2}$                       & $\frac{q v^3 \left(\tilde{f}_{\frac{5}{2}}^{\text{as1}}-f_2\right)}{b}$ \\ \hline
$\partial_\phi \psi_{\rm reg, diss}^{\eta^5, \epsilon^1}$                       & $\frac{q v^5 \left(150 (\alpha +2) f_{\frac{5}{2}}-(110 \alpha +217)
   f_{\frac{3}{2}}\right)}{30 b}$ \\  \hline
$\partial_\phi \psi_{\rm reg, diss}^{\eta^5, \epsilon^2}$                       & $\frac{q v^5 \left(3 (100 \alpha +187) \tilde{f}_{\frac{5}{2}}^{\text{as1}}-750
   (\alpha +2) \tilde{f}_{\frac{7}{2}}^{\text{as1}}-(560 \alpha +1111) f_2+750
   (\alpha +2) f_3\right)}{30 b}$ \\  \hline
$\partial_\phi \psi_{\rm reg, diss}^{\eta^6, \epsilon^2}$                       & $\frac{4 (\alpha +2) q v^6 \left(3 \tilde{f}_{\frac{5}{2}}^{\text{as1}}+f_1-3
   f_2\right)}{3 b}$ \\  \hline
\end{tabular}
\end{ruledtabular}
\end{table}

\begin{table}[]
\caption{PM-PN expansion of   $F_t$, $F_r$, $F_\phi$, later split in its conservative and dissipative parts.}\label{F_t}
\begin{ruledtabular}
\begin{tabular}{|c|l|}
\hline
$F_t^{\eta^4, \epsilon^1}$                       & $\frac{\left(3 f_{\frac{5}{2}}-2 f_{\frac{3}{2}}\right) q^2 v^4}{3 b^2}$ \\ \hline
$F_t^{\eta^4, \epsilon^2}$                       & $\frac{q^2 v^4 \left(6 \tilde{f}_{\frac{5}{2}}^{\text{as1}}-15
   \tilde{f}_{\frac{7}{2}}^{\text{as1}}-10 f_2+15 f_3\right)}{3 b^2}$ \\ \hline
$F_t^{\eta^5, \epsilon^1}$                       & $\frac{(\alpha +2) q^2 v^5 \tilde{f}_3}{b^2}$ \\  \hline
$F_t^{\eta^5, \epsilon^2}$                       & $-\frac{(\alpha +2) q^2 v^5 \left(5 f_3^{\text{as1}}-6
   \left(\tilde{f}_{\frac{7}{2}}+f_4^{\text{as1}}\right)\right)}{b^2}$ \\  \hline
$F_t^{\eta^6, \epsilon^1}$                       & $\frac{q^2 v^6 \left((20 \alpha +26) f_{\frac{3}{2}}+(121-30 \alpha )
   f_{\frac{5}{2}}-200 f_{\frac{7}{2}}\right)}{30 b^2}$ \\  \hline
$F_t^{\eta^6, \epsilon^2}$                       & $\frac{q^2 v^6 \left(-18 (20 \alpha +43) \tilde{f}_{\frac{5}{2}}^{\text{as1}}+900
   \alpha  \tilde{f}_{\frac{7}{2}}^{\text{as1}}-465
   \tilde{f}_{\frac{7}{2}}^{\text{as1}}+4200 \tilde{f}_{\frac{9}{2}}^{\text{as1}}+10
   (93 \alpha +139) f_2+25 (65-48 \alpha ) f_3-4200 f_4\right)}{90 b^2}$ \\  \hline
\hline
$F_r^{\eta^3, \epsilon^1}$                       & $\frac{2 q^2 v^3 \tilde{f}_2}{3 b^2}$ \\ \hline
$F_r^{\eta^3, \epsilon^2}$                       & $\frac{2 q^2 v^3 \left(4 \left(\tilde{f}_{\frac{5}{2}}+f_3^{\text{as1}}\right)-3
   f_2^{\text{as1}}\right)}{3 b^2}$ \\ \hline
$F_r^{\eta^5, \epsilon^1}$                       & $\frac{q^2 v^5 \left(5 (6 \alpha +1) \tilde{f}_3-(10 \alpha +13)
   \tilde{f}_2\right)}{15 b^2}$ \\  \hline
$F_r^{\eta^5, \epsilon^2}$                       & $\frac{q^2 v^5 \left(-405 \alpha  \tilde{f}_{\frac{5}{2}}+90 (6 \alpha +1)
   \tilde{f}_{\frac{7}{2}}-566 \tilde{f}_{\frac{5}{2}}+9 (20 \alpha +43)
   f_2^{\text{as1}}-3 (230 \alpha +197) f_3^{\text{as1}}+540 \alpha 
   f_4^{\text{as1}}+90 f_4^{\text{as1}}\right)}{45 b^2}$ \\  \hline
$F_r^{\eta^6, \epsilon^1}$                       & $-\frac{(\alpha +2) \left(2 f_{\frac{3}{2}}-16 f_{\frac{5}{2}}+17
   f_{\frac{7}{2}}\right) q^2 v^6}{3 b^2}$ \\  \hline
$F_r^{\eta^6, \epsilon^2}$                       & $\frac{q^2 v^6}{192 b^2} \Big[64 (\alpha +2) \left(-80 \tilde{f}_{\frac{7}{2}}^{\text{as1}}+119
   \tilde{f}_{\frac{9}{2}}^{\text{as1}}+6
   \left(\tilde{f}_{\frac{5}{2}}^{\text{as1}}+3
   \left(\tilde{f}_{\frac{5}{2}}+f_3^{\ln }\right)+3 f_3^{\text{as1}}-2 f_2^{\ln
   }\right)-12 f_2^{\text{as1}}-119 f_4\right)$ \\  
   & $+2 \left(9 \pi ^2 (\alpha  (\alpha
   +4)+7)-32 (\alpha +2) (6 \alpha +25)\right) f_2+\left(64 (\alpha +2) (9 \alpha
   +130)-27 \pi ^2 (\alpha  (\alpha +4)+7)\right) f_3$ \\
   & $+384 (\alpha +2) \left(2 f_2-3
   f_3\right) \ln (\text{$\sigma $r} v)\Big]$ \\\hline
\hline
$F_\phi^{\eta^3, \epsilon^1}$                       & $-\frac{f_{\frac{3}{2}} q^2 v^3}{3 b}$ \\ \hline
$F_\phi^{\eta^3, \epsilon^2}$                       & $\frac{q^2 v^3 \left(\tilde{f}_{\frac{5}{2}}^{\text{as1}}-f_2\right)}{b}$ \\ \hline
$F_\phi^{\eta^4, \epsilon^1}$                       & $-\frac{(\alpha +2) q^2 v^4 \tilde{f}_2}{b}$ \\  \hline
$F_\phi^{\eta^4, \epsilon^2}$                       & $\frac{(\alpha +2) q^2 v^4 \left(3 f_2^{\text{as1}}-4
   \left(\tilde{f}_{\frac{5}{2}}+f_3^{\text{as1}}\right)\right)}{b}$ \\  \hline
$F_\phi^{\eta^5, \epsilon^1}$                       & $\frac{q^2 v^5 \left(30 (2 \alpha +7) f_{\frac{5}{2}}-(50 \alpha +157)
   f_{\frac{3}{2}}\right)}{30 b}$ \\  \hline
$F_\phi^{\eta^5, \epsilon^2}$                       & $\frac{q^2 v^5 \left(3 \left((40 \alpha +127) \tilde{f}_{\frac{5}{2}}^{\text{as1}}-50
   (2 \alpha +7) \tilde{f}_{\frac{7}{2}}^{\text{as1}}+50 (2 \alpha +7)
   f_3\right)-(230 \alpha +781) f_2\right)}{30 b}$ \\  \hline
$F_\phi^{\eta^6, \epsilon^1}$                       & $-\frac{(\alpha +2) q^2 v^6 \left(13 \tilde{f}_2-4 \tilde{f}_3\right)}{6 b}$ \\\hline
$F_\phi^{\eta^6, \epsilon^2}$                       & $\frac{q^2 v^6}{96 b} \Big[\alpha  \left(384
   \left(\tilde{f}_{\frac{5}{2}}^{\text{as1}}+\tilde{f}_{\frac{7}{2}}+\tilde{f}_{\frac{5}{2}}^{\ln }\right)-\left(-288 \alpha +9 \pi ^2 (\alpha +4)+544\right)
   \tilde{f}_{\frac{5}{2}}+128 \tilde{f}_{\frac{3}{2}}-48 (6 \alpha +17)
   f_2^{\text{as1}}\right.$\\
&$\left.+384 (\alpha +2) f_3^{\text{as1}}+384 f_4^{\text{as1}}\right)$ \\
   & $+768
   \left(\tilde{f}_{\frac{5}{2}}^{\text{as1}}+\tilde{f}_{\frac{7}{2}}+\tilde{f}_{\frac{5}{2}}^{\ln }\right)-384 (\alpha +2) \tilde{f}_{\frac{5}{2}} \ln
   (\sigma_\phi v)+256 \tilde{f}_{\frac{3}{2}}-7 \left(320+9 \pi ^2\right)
   \tilde{f}_{\frac{5}{2}}-480 f_2^{\text{as1}}+768 f_4^{\text{as1}}-384 (\alpha +2)
   f_2$\\
   & $+128 (\alpha +2) f_1\Big]$\\\hline
\end{tabular}
\end{ruledtabular}
\end{table}

\end{widetext}
All the integrals here have been computed by using a cut-off regularization. The latter procedure (on purpose) is still traceable, because of the presence of the various dimensionless scales, like $\sigma_r$ and $\sigma_\phi$. All these scales do not enter the final results.

\subsection{Energy and angular momentum losses}

We can now evaluate the energy and angular momentum losses as  $F_\alpha = dP_\alpha/d\tau=\mu a(u)_\alpha$ with $\mu$ the mass of the scalar particle, $u$ its four velocity, $P=\mu u$ the four momentum, $a(u)$ its four acceleration and $F$ given in Eq. \eqref{SF_def},
\bea
\Delta E&=&\int_{-\infty}^\infty dt F_t^{\rm reg}\,,\nonumber\\
\Delta J&=&\int_{-\infty}^\infty dt F_\phi^{\rm reg}\,.
\eea
We find
\bea
\Delta E&=&\frac{\pi  \eta ^3 q^2 r_s^2}{24 b^3
   v}+\eta ^5 r_s^2 \left(\frac{\pi  \alpha  q^2 v}{16
   b^3}+\frac{31 \pi  q^2 v}{720
   b^3}\right) \nonumber\\
&=& \Delta E^{\alpha^0}+\alpha \Delta E^{\alpha^1}\,,
\eea
and
\bea
\Delta J&=&\eta ^2 \left(-\frac{\pi  q^2 r_s^2}{12 b^2
   v^2}-\frac{q^2 r_s}{3 b}\right)+\nonumber\\
   &+&\eta ^4 \left(r_s^2 \left(-\frac{\pi  \alpha 
   q^2}{24 b^2}-\frac{\pi  q^2}{10
   b^2}\right)+r_s \left(-\frac{\alpha  q^2
   v^2}{3 b}\right.\right.\nonumber\\
   &-&\left.\left.\frac{17 q^2 v^2}{30 b}\right)\right)\nonumber\\
&=& \Delta J^{\alpha^0}+\alpha \Delta J^{\alpha^1}\,.
\eea
Notice the $\alpha$-modifications to the Schwarzschild-like results
\bea
\Delta E^{\alpha^1}&=&  \eta ^5 r_s^2 \frac{\pi  q^2 v}{16b^3}\,,\nonumber\\
\Delta J^{\alpha^1}&=& -\eta ^4 \frac{r_sq^2}{3b}\left(  \frac{\pi}{8}\frac{r_s}{ b}    + v^2 \right)\,.
\eea
These terms (an original accomplishment of this work) represent  deviations from the corresponding Schwarzschild-like behaviors~\footnote{The TS modifications are not simply Schwarzschild plus $\alpha$-terms, due to the fact that in the TS spacetime $r>r_b>r_s$: these conditions have been taken into account when imposing boundary conditions to the radial equation.}, and should be compared with numerical simulation data as soon as they will be available.
 
\section{Discussion}

In the present investigation we have discussed another application of self-force in the spacetime of a Top Star reduced to $D=4$, i.e. taking advantage of a previous study and neglecting emission of rapidly decaying massive KK modes.
Indeed, in a previous investigation we considered the case of massless scalar waves -- as a proxy of gravitational waves -- sourced by a particle moving on a circular orbit, and in the present paper we approached the more involved situation of a massless scalar field sourced by a massive particle moving along an unbound orbit. 
We have been working in a doubly expanded PM-PN framework ($\epsilon-\eta$ expansion above), which recent studies in black hole spacetimes have proven to be very convenient and the novel approach based on $N=2$ supersymmetric quiver gauge theories and the AGT correspondence \cite{Alday:2009aq} has rendered very natural \cite{Aminov:2020yma, Bianchi:2021xpr, Bonelli:2021uvf, Bianchi:2021mft, Fioravanti:2021dce, Bianchi:2022qph, Consoli:2022eey, Bonelli:2022ten, Gregori:2022xks, 
Fucito:2023afe, Bautista:2023sdf, Bianchi:2023rlt, Aminov:2023jve, DiRusso:2024hmd, Fucito:2024wlg, Bianchi:2024mlq, Cipriani:2024ygw, Bena:2024hoh}.

We succeeded in fully reconstructing the field along the source's world line both in the time domain and in the frequency domain.
Clearly, in the present case the frequency spectrum is continuous, whereas for circular orbits one has a single frequency only. This constitutes a major complication.
As a by-product our study shows how to encompass the main difficulties underlying a similar treatment of the gravitational field itself (metric and curvature perturbations) to be attempted in the future.
We learned the structure of the time domain field which, besides terms of the type $(1+T^2)^{-n}$ (with $T=v t/b$ a dimensionless temporal variable),   
involves terms like ${\rm arcsinh}^m(T)(1+T^2)^{-n}$, ${\rm arctan}^m(T)(1+T^2)^{-n}$,  ${\rm arctan}^q(T){\rm arcsinh}^m(T)(1+T^2)^{-n}$, etc. which strongly limit (or even forbid) an explicit analytic conversion into the frequency domain. 
Indeed, rather involved Meijer $G$ functions appear, too.
We plan to accomplish the task of reaching high PM-PN results for the radiative losses as well as to compute the conservative and radiative contributions to the scattering angle in future works. 
In the meantime, we are also planning comparisons with numerical relativity results, which would validate these results and motivate additional analytic efforts in going beyond the scope of the present  PM-PN double expansion.

\appendix

\section{Hadamard's partie finie of divergent integrals}
\label{PF_recalls}

Hadamard regularization (also called Hadamard finite part or Hadamard's partie finie) is a method of regularizing divergent integrals by dropping some divergent terms and keeping the finite part. The definition is the following
\begin{eqnarray}
\label{orig_def}
{\rm Pf}_T \int _0^{+\infty} \frac{dv}{v}g(v)&\equiv& \int_0^a \frac{dv}{v}(g(v)-g(0))\nonumber\\
&+&\int_a^\infty \frac{dv}{v}g(v)+g(0) \ln \frac{a}{T}\,,\qquad
\end{eqnarray}
where $T$ is a \lq\lq scale"  and $a$ is an arbitrary constant value.
Practically, the two divergent integrals 
\beq
g(0)\int_0^a \frac{dv}{v}\,,\qquad g(0)\int_T^\infty \frac{dv}{v}\,,
\eeq
(the first diverging in $v\to 0$ and the second in $v\to \infty$) 
are excluded.

The definition implies that the Pf integral does not depend on $a$, in the sense that
\beq
\frac{d}{da} \left[ {\rm Pf}_T \int _0^{+\infty} \frac{dv}{v}g(v) \right] =0\,.
\eeq
Integral equations containing Hadamard finite part integrals (with the integrand unknown) are termed {\it hypersingular integral equations}. 
As an example consider the integral (see e.g., Eq. 5.8 of Ref.  \cite{Damour:2014jta})
\beq
{\rm Pf}_T \int _0^{+\infty} \frac{d\tau}{\tau} \cos(\omega \tau)=-\gamma -\ln (|\omega | T)\,.
\eeq
Using $a=1/\omega$ and changing the integration variable as $x=\omega \tau$ (the sign of $\omega$ being inessential) we find
\bea
\label{pf_cos}
{\rm Pf}_T \int _0^{+\infty} \frac{d\tau}{\tau} \cos(\omega \tau)&=&\underbrace{\int_0^1  \frac{dx}{x} [\cos(x)-1]}_{-\gamma+{\rm Ci}(1)}\nonumber\\
&+&\underbrace{\int_1^\infty \frac{\cos(x)}{x}dx}_{-{\rm Ci}(1)}-\ln (|\omega| T)\nonumber\\
&=&-\gamma -\ln (|\omega | T)\nonumber\\
&=&-\ln(e^{\gamma}|\omega|T)\,.
\eea
Here,
\beq
{\rm Ci}(x)=\gamma + \ln(x) + \int_0^x  \frac{ \cos(t)-1 }{t}dt\,.
\eeq
Eq. 4.2 of Ref.  \cite{Damour:2014jta} is consistent with the definition 
\eqref{orig_def} with $a=T$, 
\beq
{\rm Pf}_T\int_0^\infty \frac{dv}{v}g(v)=\int_0^T \frac{dv}{v}[g(v)-g(0)]+\int_T^\infty \frac{dv}{v}g(v)\,.
\eeq

\section{Evaluating useful integrals/Fourier transforms}
\label{appB}

\begin{widetext}

It is often sufficient to regularize divergent integrals by using a (single) cut-off $\epsilon>0$. For example, 
\beq
\label{cut-off_integ}
{\mathcal I}=\int_{-\infty}^t dt' \frac{F(t,t')}{t-t'}=\lim_{\epsilon \to 0}\int_{-\frac{1}{\epsilon}}^{t-\epsilon} dt' \frac{F(t,t')}{t-t'}\,.
\eeq
The evaluated integral is written as a Laurent series of $\epsilon$ plus logs
\beq
{\mathcal I}=\sum_{k=-\infty}^\infty c_k \epsilon^k + \sum_{k=1}^\infty d_k\ln^k \epsilon\,,
\eeq  
and its finite part results in $c_0$ plus $\ln(\epsilon)$ (namely, powers $\ln^n(\epsilon)$ with $n>1$ cancel). 
In fact, if one adds and subtracts $F(t,t)$ in Eq. \eqref{cut-off_integ} one finds
\beq
\label{cut-off_integ2}
{\mathcal I}=\int_{-\infty}^t dt' \frac{F(t,t')-F(t,t)}{t-t'}+F(t,t)\int_{-\infty}^t \frac{dt'}{t-t'}\,.
\eeq
Here, the first integral is no more singular at $t'=t$ and from the second one only logs to the first power may arise.

Relevant integrals for the present study (and for those which will reach higher PN approximations) are of the type
\beq
\label{gen_omega}
F_n(t)=\int \frac{d\omega}{2\pi} e^{-i\omega t}\ln^n(\omega)\,,
\eeq
and evaluate as follows
\bea
\label{gen_omega1}
F_n(t)&=&\frac{\mathcal{A}_n(t)}{t}H(t)+B_n \delta(t)\nonumber\\
&=&\frac{A_n+A_n^{\ln}\ln (t)+A_n^{\ln^2}\ln^2 (t)+\ldots}{t}H(t)+B_n \delta(t)\nonumber\\
&=&\left(A_n \frac{d}{dt}\ln t+ \frac{A_n^{\ln}}2 \frac{d}{dt}\ln^2 t+\frac{A_n^{\ln^2}}{3}\frac{d}{dt}\ln^3 t+\ldots \right)H(t)+B_n \delta(t)\,,
\eea
that is
\bea
\label{B5}
F_1(t)&=& -\frac{1}{t}H(t)+\left(-\gamma +i\frac{\pi}{2} \right)\delta(t)\,,\nonumber\\
F_2(t)&=& \frac{ 2\gamma  -i\pi + 2\ln t}{t} H(t)+\left(\gamma^2 -i\pi\gamma - \frac{1}{3}\pi^2  \right)\delta(t)\,,\nonumber\\
F_3(t)&=&\frac{-12\gamma^2 + 12i\gamma \pi+  5 \pi^2+ (-24\gamma + 12i\pi)\ln(t) - 12\ln^2(t)}{4 t} H(t)\nonumber\\
&+&\left(-\gamma^3 + \frac32 i \pi\gamma^2  + \pi^2\gamma-\frac{i\pi^3}{4}  - 2 \zeta (3)\right)\delta(t)\,,\nonumber\\
F_4(t)&=&\frac{1}{2t}[8\gamma^3 - 12i\gamma^2\pi- 10\gamma \pi^2 + 3 i\pi^3 + 2 (12 \gamma^2 - 12i \gamma \pi - 5 \pi^2) \ln (t)\nonumber\\
&+&  12 (2 \gamma - i\pi )\ln^2 (t) + 8 \ln^3(t) + 16 \zeta(3)]H(t)\nonumber\\
&+& \left(\gamma^4 - 2i\pi\gamma^3  - 2\gamma^2 \pi^2 + 
 \gamma(i\pi^3+8\zeta(3))+\frac{3\pi^4}{20}-4i\pi \zeta(3) \right)\delta (t)\,.
\eea
In general, by using algebraic manipulator systems, we can write
\bea
\mathcal{A}_n(t)&=&=\frac{i}{2\pi}\sum_{k_1=1}^n \binom{n}{k_1}\Big[B_{n-k_1}-\frac{1}{2}\sum_{k_2=1}^{n-k_1}\binom{n-k_1}{k_2}(i\pi)^{k_2}\left(\frac{d^{n-k_1-k_2}}{dk^{n-k_1-k_2}}\Gamma(1+k)\right) \left.\right|_{k{=}0}\Big]\nonumber\\
&\times&\Big[(-1)^{k_1+1}\left(\frac{i\pi}{2}+\ln(t)\right)^{k_1}+\left(\frac{3i\pi}{2}-\ln(t)\right)^{k_1}\Big]\,.
\eea

\begin{table}[]
\caption{\label{tabFTlog_om} List of the coefficients appearing in Eq. \eqref{gen_omega}.}
\begin{ruledtabular}
\begin{tabular}{|c|c|c|c|c|c|}
\hline
n & $A_n$                                                                                     & $A_n^{\ln}$           & $A_n^{\ln^2}$ & $A_n^{\ln^3}$ & $B_n$                                         \\ \hline
1 & -1                                                                                        & 0                     & 0             & 0             & $ -\gamma +i\frac{\pi}{2}$                    \\ \hline
2 & $-2B_1$                                                                                   & 2                     & 0             & 0             & $ B_1^2-\frac{\zeta (2)}{2}$                  \\ \hline
3 & $-3B_2-3i\pi B_1$                                                                         & $6B_1$                & -3            & 0             & $B_1^3-\frac{3}{2} B_1 \zeta (2)-2 \zeta (3)$ \\
  & $-\frac{5\pi^2}{4}-3i \gamma \pi$                                                         &                       &               &               &                                               \\ \hline
4 & $-4 B_3+\pi  (7 \pi  B_1-6 i B_2$                                                         & $12B_2+\pi(12i\gamma$ & $-12B_1$      & 4             & $B_1^4 -3 B_1^2 \zeta (2)-8 B_1 \zeta (3)$    \\
  & \begin{tabular}[c]{@{}c@{}}$-5 i \pi ^2+12   \gamma  \pi +6 i \gamma ^2)$\end{tabular} & $+5\pi+12iB_1)$       &               &               & $-\frac{27 \zeta (4)}{8}\zeta(4)$             \\ \hline
\end{tabular}
\end{ruledtabular}
\end{table}

Finally, 
\bea
B_n&{=}&\frac{1}{2}\sum_{k_1=0}^n\binom{n}{k_1}\left. \left(\frac{d^{n-k_1}}{dk^{n-k_1}}\Gamma(1+k)\right) \right|_{k{=}0}
\begin{array}{cc}
 \Bigg\{ & 
\begin{array}{cc}
 (-1)^{k_1}+1 & k_1=0 \\
 (i\pi)^{k_1} & k_1>0 \\
\end{array}
 \\
\end{array}
\Bigg]\,.
\eea

Following Ref. \cite{Bini:2024icd} we can prove the result for the integral $F_2(t)$. To this end let us   
consider  the deformed integral
\be
g_\epsilon^{(2)}=\int_{-\infty}^{\infty}\frac{d\omega}{2\pi}e^{-i \omega  t-\epsilon|\omega|}\ln^2(\omega)\,,
\ee
which can be exactly integrated
\bea\label{gepsilon}
g_\epsilon^{(2)}(t)&=&-\frac{i \pi  t}{2 \left(t^2+\epsilon ^2\right)}+\frac{\gamma  t}{t^2+\epsilon
   ^2}-\frac{\pi  \epsilon }{3 \left(t^2+\epsilon ^2\right)}+\frac{\gamma ^2
   \epsilon }{\pi  \left(t^2+\epsilon ^2\right)}-\frac{i \gamma  \epsilon
   }{t^2+\epsilon ^2}+\frac{\epsilon  \ln ^2\left(t^2+\epsilon ^2\right)}{4 \pi 
   (t^2+  \epsilon ^2)}+\frac{\pi  t \ln \left(t^2+\epsilon ^2\right)}{2 \pi 
   (t^2+\epsilon ^2)}\nonumber\\
   &-&\frac{i \pi  \epsilon  \ln \left(t^2+\epsilon
   ^2\right)}{2 \pi  (t^2+  \epsilon ^2)}+\frac{2 \gamma  \epsilon  \ln
   \left(t^2+\epsilon ^2\right)}{2 \pi  (t^2+  \epsilon ^2)}-\frac{\epsilon \,
   {\rm arctan}\left(\frac{t}{\epsilon }\right)^2}{\pi  (t^2+  \epsilon
   ^2)}-\frac{i \pi  t\, {\rm arctan}\left(\frac{t}{\epsilon }\right)}{\pi  (t^2+ 
   \epsilon ^2)}+\frac{2 \gamma  t \, {\rm arctan}\left(\frac{t}{\epsilon }\right)}{\pi(
    t^2+  \epsilon ^2)}\nonumber\\
    &-&\frac{\pi  \epsilon  \, {\rm arctan}\left(\frac{t}{\epsilon
   }\right)}{\pi  (t^2+  \epsilon ^2)}+\frac{t \ln \left(t^2+\epsilon ^2\right)
   {\rm arctan}\left(\frac{t}{\epsilon }\right)}{\pi  (t^2+  \epsilon ^2)}\,.
\eea
From the previous relation we can isolate the terms that reproduce the Dirac delta (i.e., define a Dirac delta approximant)
\beq
g_{\epsilon,\delta}^{(2)}(t)=\left(-\frac{\pi^2}{3}+ \gamma ^2- i\pi \gamma \right) \frac{\epsilon}{\pi \left(t^2+\epsilon ^2\right)}\,.
\eeq
Applying $g_{\epsilon,\delta}^{(2)}$ to a test function $\phi(t)$ and taking the limit  $\epsilon\to 0$ gives
\beq
\lim_{\epsilon\to 0}\int_{-\infty}^\infty dt g_{\epsilon,\delta}^{(2)}(t) \phi(t)=\left(\gamma^2-i \pi \gamma-\frac{\pi^2}{3}\right)\phi(0)\,,
\eeq
implying that 
\beq\label{gdelta}
g_{\epsilon,\delta}^{(2)}=\left(\gamma^2-i \pi \gamma-\frac{\pi^2}{3}\right)\delta(t)
\eeq
in the sense of the distributions.
All the remaining terms in \eqref{gepsilon}, $g_{\epsilon,{\rm H}}^{(2)}(t)=g_\epsilon^{(2)}(t)-g_{\epsilon,\delta}^{(2)}(t)$,  in the limit $\epsilon\to 0$  give instead\footnote{The Heaviside part of these integrals have been checked numerically for several specified values of $t$.}
\be
\label{gheav}
\lim_{\epsilon=0} g_{\epsilon,{\rm H}}^{(2)}(t)=\frac{H(t)(2\gamma-i \pi+2\ln(t))}{t}\,.
\ee
Eqs. \eqref{gdelta} and \eqref{gheav} reproduce the same result appearing in Eq. \eqref{B5} and Table \ref{tabFTlog_om}. Note that Ref. \cite{Bini:2024icd} (see e.g. Eq. (A18) and related equations there) shows a fully distributional form of these integrals which one could clearly also be repeated here.

Similarly, regarding $F_3(t)$, we can write
\be
g_\epsilon^{(3)}=\int_{-\infty}^{\infty}\frac{d\omega}{2\pi}e^{-i \omega  t-\epsilon|\omega|}\ln^3(\omega)\,,
\ee
which leads to
\bea
g_\epsilon^{(3)}&=&\frac{i t \log ^3(i t+\epsilon )}{\pi  \left(t^2+\epsilon ^2\right)}-\frac{3 i t \log \left(t^2+\epsilon
   ^2\right) \log ^2(i t+\epsilon )}{2 \pi  \left(t^2+\epsilon ^2\right)}-\frac{3 \epsilon  \log
   \left(t^2+\epsilon ^2\right) \log ^2(i t+\epsilon )}{2 \pi  \left(t^2+\epsilon ^2\right)}-\frac{3 t \log ^2(i
   t+\epsilon )}{2 \left(t^2+\epsilon ^2\right)}-\frac{3 \gamma  \epsilon  \log ^2(i t+\epsilon )}{\pi 
   \left(t^2+\epsilon ^2\right)}\nonumber\\
   &+&\frac{3 i \epsilon  \log ^2(i t+\epsilon )}{2 \left(t^2+\epsilon
   ^2\right)}+\frac{3 i t \log ^2\left(t^2+\epsilon ^2\right) \log (i t+\epsilon )}{2 \pi  \left(t^2+\epsilon
   ^2\right)}+\frac{3 \epsilon  \log ^2\left(t^2+\epsilon ^2\right) \log (i t+\epsilon )}{2 \pi 
   \left(t^2+\epsilon ^2\right)}+\frac{3 i \gamma  t \log \left(t^2+\epsilon ^2\right) \log (i t+\epsilon )}{\pi
    \left(t^2+\epsilon ^2\right)}\nonumber\\
    &+&\frac{3 t \log \left(t^2+\epsilon ^2\right) \log (i t+\epsilon )}{t^2+\epsilon
   ^2}+\frac{3 \gamma  \epsilon  \log \left(t^2+\epsilon ^2\right) \log (i t+\epsilon )}{\pi  \left(t^2+\epsilon
   ^2\right)}-\frac{3 i \epsilon  \log \left(t^2+\epsilon ^2\right) \log (i t+\epsilon )}{t^2+\epsilon
   ^2}-\frac{i \pi  t \log (i t+\epsilon )}{t^2+\epsilon ^2}\nonumber\\
   &+&\frac{3 i \gamma ^2 t \log (i t+\epsilon )}{\pi 
   \left(t^2+\epsilon ^2\right)}+\frac{3 \gamma  t \log (i t+\epsilon )}{t^2+\epsilon ^2}-\frac{3 \pi  \epsilon 
   \log (i t+\epsilon )}{2 \left(t^2+\epsilon ^2\right)}-\frac{3 i \gamma  \epsilon  \log (i t+\epsilon
   )}{t^2+\epsilon ^2}-\frac{i t \log ^3\left(t^2+\epsilon ^2\right)}{2 \pi  \left(t^2+\epsilon
   ^2\right)}-\frac{\epsilon  \log ^3\left(t^2+\epsilon ^2\right)}{2 \pi  \left(t^2+\epsilon ^2\right)}\nonumber\\
   &-&\frac{3
   i \gamma  t \log ^2\left(t^2+\epsilon ^2\right)}{2 \pi  \left(t^2+\epsilon ^2\right)}-\frac{3 t \log
   ^2\left(t^2+\epsilon ^2\right)}{2 \left(t^2+\epsilon ^2\right)}-\frac{3 \gamma  \epsilon  \log
   ^2\left(t^2+\epsilon ^2\right)}{2 \pi  \left(t^2+\epsilon ^2\right)}+\frac{3 i \epsilon  \log
   ^2\left(t^2+\epsilon ^2\right)}{2 \left(t^2+\epsilon ^2\right)}+\frac{5 i \pi  t \log \left(t^2+\epsilon
   ^2\right)}{4 \left(t^2+\epsilon ^2\right)}\nonumber\\
   &-&\frac{3 i \gamma ^2 t \log \left(t^2+\epsilon ^2\right)}{2 \pi 
   \left(t^2+\epsilon ^2\right)}-\frac{3 \gamma  t \log \left(t^2+\epsilon ^2\right)}{t^2+\epsilon ^2}+\frac{5
   \pi  \epsilon  \log \left(t^2+\epsilon ^2\right)}{4 \left(t^2+\epsilon ^2\right)}-\frac{3 \gamma ^2 \epsilon 
   \log \left(t^2+\epsilon ^2\right)}{2 \pi  \left(t^2+\epsilon ^2\right)}+\frac{3 i \gamma  \epsilon  \log
   \left(t^2+\epsilon ^2\right)}{t^2+\epsilon ^2}\nonumber\\
   &+&\frac{\pi ^2 t}{4 \left(t^2+\epsilon ^2\right)}+\frac{3 i
   \gamma  \pi  t}{2 \left(t^2+\epsilon ^2\right)}-\frac{3 \gamma ^2 t}{2 \left(t^2+\epsilon ^2\right)}-\frac{i
   \pi ^2 \epsilon }{4 \left(t^2+\epsilon ^2\right)}+\frac{\gamma  \pi  \epsilon }{t^2+\epsilon ^2}-\frac{\gamma
   ^3 \epsilon }{\pi  \left(t^2+\epsilon ^2\right)}+\frac{3 i \gamma ^2 \epsilon }{2 \left(t^2+\epsilon
   ^2\right)}-\frac{2 \epsilon  \zeta (3)}{\pi  \left(t^2+\epsilon ^2\right)}\,.
\eea
As before, from the latter we can easily distinguish the Dirac delta approximant from the rest, which will reconstruct the Heaviside function in the $\epsilon\to 0$ limit.

In order to see how the contribution coming from the Heaviside step function to $g_{\epsilon=0}$ arise let us perform the contour integral on the path showed in Fig. \ref{contour} as in Ref. \cite{Racine:2008kj}. It is easy to show that the contributions coming from the arcs $C_2$ and $C_4$ vanish in the limits $R_1\to 0$ and $R_2\to\infty$. Since in the chosen path the function has no singularities the contour integral on all the path is zero, so we are left with
\be
\int_{C_1} \frac{d\omega}{2\pi} e^{-i \omega t}\ln^2(\omega)=-\int_{C_3} \frac{d\omega}{2\pi}e^{-i \omega t}\ln^2(\omega)\,,
\ee
so in the limit $R_1\to 0$ and $R_2\to \infty$ the final result is
\be\label{racinne}
\int_0^\infty \frac{d\omega}{2\pi} e^{-i \omega t}\ln^2(\omega)=\frac{-12 i \ln ^2(t)+12 (\pi -2 i \gamma) \ln (t)+i \pi ^2+12 \gamma   \pi -12
   i \gamma^2}{24 \pi  t}\,.
\ee

\begin{figure}
\includegraphics[scale=0.55]{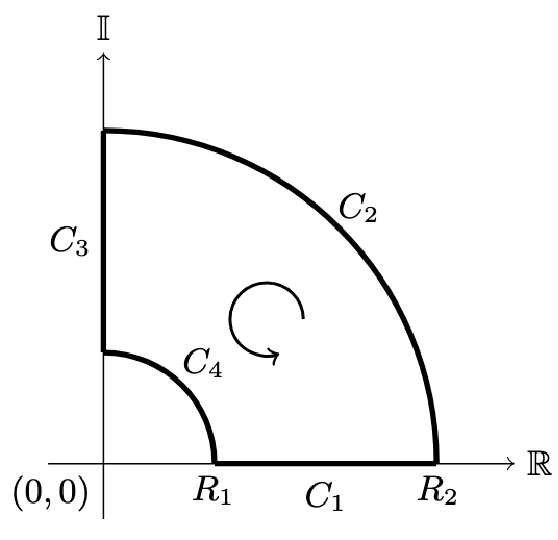}
\caption{\label{contour} Contour used to integrate $g_{\epsilon=0}$ as defined in \eqref{gheav}.}
\end{figure}

A similar procedure can be applied iteratively for the Fourier transform of various integrals $F_n(t)$. The coefficient of the Heaviside theta appearing in the second equation in \eqref{B5} can be straightforwardly reproduced by \eqref{racinne} by simply changing $t\to -t$ and introducing the new variable $\omega\to-\omega$ in order to add the integral from $-\infty$ to $0$.

\end{widetext}

\section*{Acknowledgments}

We  thank  A.~Argenzio, A.~Cipriani, G.~Dibitetto, F.~Fucito, A.~Geralico,  J.~F.~Morales and A.~Ruiperez
Vicente for useful discussions and comments. 
D.~B. thanks D.~Usseglio for informative discussions on the Fourier transforms of log terms and acknowledges sponsorship of the Italian Gruppo Nazionale per la Fisica Matematica
(GNFM) of the Istituto Nazionale di Alta Matematica (INDAM).
M.~B. and G.~D.~R. thank the MIUR PRIN contract 2020KR4KN2 \lq\lq String Theory as a bridge
between Gauge Theories and Quantum Gravity'' and the INFN project ST\&FI \lq\lq String Theory and
Fundamental Interactions'' for partial support.

\end{document}